\documentclass[twocolumn]{revtex4-1}
\usepackage{graphicx}
\usepackage{dcolumn}
\usepackage{amsmath}

\begin{document}

\title{Green's functions of Nambu-Goldstone modes and Higgs modes 
in superconductors}

\author{Takashi Yanagisawa}

\affiliation{Electronics and Photonics Research Institute,
National Institute of Advanced Industrial Science and Technology (AIST),
Tsukuba Central 2, 1-1-1 Umezono, Tsukuba 305-8568, Japan
}

\begin{abstract}
We examine fundamental properties of Green's functions of Nambu-Goldstone and
Higgs modes in superconductors with multiple order parameters. 
Nambu-Goldstone and Higgs modes are determined once the
symmetry of the system and that of the order parameters are
specified.
Multiple Nambu-Goldstone modes and Higgs modes exist when
we have multiple order parameters.
The Nambu-Goldstone Green function $D(\omega,{\bf q})$ has the form
$1/(gN(0))^2\cdot (2\Delta)^2/(\omega^2-c_s^2{\bf q}^2)$
with the coupling constant $g$ and $c_s=v_F/\sqrt{3}$ for small $\omega$
and ${\bf q}$, with a pole at
$\omega=0$ and ${\bf q}=0$ indicating the existence of a massless mode.
It is shown, based on the Ward-Takahashi identity, that the massless
mode remains massless in the presence of intraband scattering due to
nonmagnetic and magnetic impurities.
The pole of $D(\omega,{\bf q})$, however, disappears as $\omega$ increases
as large as $2\Delta$: $\omega\sim 2\Delta$.
The Green function $H(\omega,{\bf q})$ of the Higgs mode is given by
$H(\omega,{\bf q})\propto (2\Delta)^2/((2\Delta)^2-\frac{1}{3}\omega^2
+\frac{1}{3}c_s^2{\bf q}^2)$ for small $\omega$ and ${\bf q}$.
$H(\omega,{\bf q})$ is proportional to 
$1/(gN(0))^2\cdot \Delta/\sqrt{ (2\Delta)^2+c_s^2{\bf q}^2-\omega^2}$
for $\omega\sim 2\Delta$ and $\omega < \omega({\bf q})$
where $\omega({\bf q})=\sqrt{ (2\Delta)^2+c_s^2{\bf q}^2}$.
This behavior is similar to that of the $\sigma$-particle Green
function in the Gross-Neveu model.  That is, the Higgs Green function
$H(\omega, {\bf q})$ has the same singularity as the Green function of
the $\sigma$ boson of the Gross-Neveu model.
The constant part of the action for the Higgs modes is important
since it determines the
coherence length of a superconductor.
There is the case that it has a large eigenvalue,
indicating that the large upper critical field $H_{c2}$ may be realized
in a superconductor with multiple order parameters.
\end{abstract}


\maketitle

\section{Introduction}
When global and continuous symmetries are spontaneously broken,
gapless excitation modes, called the Nambu-Goldstone (NG) bosons, exist and
govern the long-distance behaviors of the system.
The spontaneous symmetry breaking indicates that the state is not
invariant under a symmetry transformation although the Lagrangian
is invariant under this transformation.
The spontaneous symmetry breaking occurs when an asymmetric state is 
realized in a symmetric system.
When a continuous symmetry is spontaneously broken, a massless boson
appears, called the Nambu-Goldstone boson (NG boson)\cite{gol61,nam60,gol62,wei95}.
The spontaneous symmetry breaking has been studied intensively in
condensed matter physics\cite{gin50,bar57,and84,abr88,whi06,gol92}
and in field 
theory\cite{nam61,hig64,gol66,wei72,nie76,col85,bra10,wat12,hid13,oda13,yan16}.

A superconducting transition is a typical example of the spontaneous
symmetry breaking.  The Nambu-Goldstone boson and also the Higgs 
boson appear associated with this transition.
The second-order phase transition that occurs as a spontaneous 
symmetry breaking is characterized by the order parameter.
A multi-component (multi-band) superconductor has been studied as 
a generalization of the Bardeen-Cooper-Schrieffer (BCS) theory\cite{bcs}.
The study of multi-band superconductivity started from works by
Moskalenko\cite{mos59}, Suhl et al.\cite{suh59}, Peretti\cite{per62}
and Kondo\cite{kon63}.
There appear many interesting properties in multi-band
superconductors such as time-reversal symmetry 
breaking\cite{sta10,tan10a,tan10b,dia11,yan12,hu12,sta12,pla12,mai13,wil13,gan14,yer15,hil09,has09}, 
the existence of massless modes\cite{yan13,lin12,kob13,koy14,yan14,tan15},
unusual isotope effect\cite{cho09,shi09,yan09}
and the existence of fractionally quantized-flux 
vortices\cite{izy90,vol09,tan02,kup11,tan18,yan18}.
We have multiple order parameters, and thus there appear multiple
Nambu-Goldstone bosons and Higgs 
bosons\cite{yan13,lit82,cea14,pek15,cea15,yan15,koy16,yan17,ait99,mur17}.
This will result in significant excitation modes that are unique
in multi-band superconductors.
The phase-difference mode between two order parameters is sometimes
called the Leggett mode\cite{leg66}.
An effective model for the dynamics of the phase-difference mode, that is,
the sine-Gordon model has also been examined\cite{yan16,yan12,yan13,yan18b}.

The purpose of this paper is to investigate properties of
Green's functions of Nambu-Goldstone bosons (modes) and Higgs bosons in 
superconductors.
The Nambu-Goldstone mode is a phase mode of the order parameter
and the Higgs mode is a fluctuation mode of the amplitude of
the order parameter.
We have interband couplings $g_{mn}$ as well as intraband attractive
couplings $g_{nn}$ in a multi-band superconductor, where
$m$ and $n$ stand for band indices.
The matrix $g\equiv (g_{mn})$ determines the property of superconductors.
The Green functions also show dependence on the matrix $g$.
We investigate the dispersion relation of excitation gaps.

In an $N$-band superconductor, there are $N$ Nambu-Goldstone modes.
We have one gapless mode (Nambu-Goldstone mode) and the other $N-1$
modes are massive (called the Nambu-Goldstone-Leggett or Leggett modes) 
in general when there are non-zero interband couplings 
$g_{nm}$.
The Nambu-Goldstone mode is a mode described by the quasiparticle excitation
mode, namely the Green function $D(\omega,{\bf q})$ is proportional to
$D(\omega,{\bf q})\propto (2\Delta)^2/(\omega^2-c_s^2{\bf q}^2)$ with
finite residue for small $(\omega,{\bf q})$ (where $c_s=v_F/\sqrt{3}$
for the Fermi velocity $v_F$). 
$D(\omega,{\bf q})$ has, however, no singularity when
$\omega\sim 2\Delta$ (where $\Delta$ is the gap function).

When the time reversal symmetry is broken, which depends on
the matrix $g$, some of Leggett modes become gapless when $N$ is
greater than 2.
We can incorporate the effects of interaction in Green's function,
using the Ward-Takahashi identity.   
The Nambu-Goldstone mode remains gapless even with electron
scattering due to impurities.

We also examine the property of Green's function of the Higgs mode.
The kinetic term of the action of Higgs modes is dependent upon
temperature.  The Higgs action reduces to the time-dependent
Ginzburg-Landau model (TDGL) with dissipation  when the temperature is close to
the critical temperature $T_c$.
At low temperatures, instead, the action is given by the quadratic form without
dissipation.
We have $H(\omega,{\bf q})\propto (\omega^2-(2\sqrt{3}\Delta)^2-c_s^2{\bf q}^2)^{-1}$
for small $\omega$ and ${\bf q}$, based on the BCS theory.  
This has no pole when $\omega$ is small
as far as $0\le \omega\le 2\Delta$.
When $\omega\sim 2\Delta$, $H(\omega,{\bf q})$ is given by
$H(\omega,{\bf q})\propto \Delta/\sqrt{(2\Delta)^2+c_s^2{\bf q}^2-\omega^2}$.

We also mention that
the constant term of the action of Higgs modes is important since
it is related with the upper critical field $H_{c2}$.
The eigenvalue $y_H^2$ of the constant term of the action of Higgs 
bosons is enhanced extremely or is softened, 
depending on the coupling constant matrix $g$.
Because $y_H$ is proportional to the inverse of
the coherence length $\xi$, the upper critical field $H_{c2}$
scales linearly with the square of $y_H$:  $H_{c2}\propto y_H^2$.
The large eigenvalue $y_H$ indicates a possibility of the large 
critical field $H_{c2}$.

The paper is organized as follows.  In Section II, we briefly
show formulas for spontaneous symmetry breaking that are necessary
in later Sections.
In Section III, we examine the properties of Green's functions
of Nambu-Goldstone modes in superconductors. 
In Section IV, the plasma mode is investigated in the presence 
of electromagnetic scalar potential.
We discuss Green's functions of the Higgs modes in Section V.
We give a summary in last Section.

\section{Formulas for spontaneous symmetry breaking}

In this section, we give a brief formal theory on spontaneous
symmetry breaking. This can be applied to multi-band superconductivity.
Let us assume that the system is invariant under the continuous
transformation given by a compact Lie group $G$.
${\bf g}$ denotes the Lie algebra of $G$.
The elements of the basis set of ${\bf g}$ are denoted as $T_a$
($a=1,\cdots,N_G$) where $N_G$ is the dimension of $G$. 
The transformation of the fermion field $\psi$ is represented as
\begin{equation}
\psi\rightarrow e^{-i\theta T_a}\psi = \psi-i\theta T_a\psi
+O(\theta^2),
\end{equation}
where $\theta$ is an infinitesimal real parameter.
We set $\delta\psi_a= -i\theta T_a\psi$ for this transformation.
When the Lagrangian $\mathcal{L}$ or the Hamiltonian is invariant 
under the transformation $\psi\rightarrow \psi+\delta\psi_a$,
there is a conserved current.
\begin{equation}
j_a^{\mu}= \frac{\delta\mathcal{L}}{\delta(\partial_{\mu}\psi)}
\delta\psi_a.
\end{equation}
We have $\partial_{\mu}j_a^{\mu}=0$.  The conserved quantities are
given by
\begin{equation}
Q_a = \int d{\bf r}J_a^0({\bf r}),
\end{equation}
where we defined
\begin{equation}
J_a^{\mu}=j_a^{\mu}/\theta.
\end{equation}

In spontaneous symmetry breaking, the ground state loses a
part of symmetry that the Lagrangian possesses.
We introduce an infinitesimal term in the Lagrangian to consider a
spontaneous symmetry breaking:
\begin{equation}
\mathcal{L}_{SB}= \lambda\psi^{\dag}M\psi,
\end{equation}
where $\lambda$ is a real infinitesimal parameter and $M$ is 
a hermitian matrix in the basis set of the Lie algebra ${\bf g}$,
that is, $M\in \{T_a\}$.
$M$ indicates a broken symmetry and can be a linear combination
of $T_a$.  The second-order phase transition is characterized
by the order parameter, where the order parameter is gien by
the expectation value of the symmetry breaking term:
\begin{equation}
\Delta = \langle\psi^{\dag}M\psi\rangle.
\end{equation}
The spontaneous symmetry breaking occurs when $\Delta$ does not
vanish, $\Delta\neq 0$, in the limit $\lambda\rightarrow 0$.
Under the transformation $\psi\rightarrow \psi-i\theta T_a\psi$,
$\mathcal{L}_{SB}$ is transformed to 
$\mathcal{L}_{SB}+\delta\mathcal{L}_{SB}$ where
\begin{equation}
\delta\mathcal{L}_{SB} = i\theta\lambda\psi^{\dag}[T_a,M]\psi.
\end{equation}
Then, the corresponding current is not conserved:
\begin{equation}
\partial_{\mu}J_a^{\mu} = \delta\mathcal{L}_{SB}
= i\lambda\psi^{\dag}[T_a,M]\psi.
\end{equation}
The Nambu-Goldstone boson is given by
\begin{equation}
\pi_a = i\psi^{\dag}[T_a,M]\psi.
\end{equation}
$\pi_a$ indeed indicates a massless boson, which is shown by
evaluating the Green's function,
\begin{equation}
D_{aa}(x-y) = -i\langle T\pi_a(x)\pi_b(y)\rangle.
\end{equation}
The Fourier transform of $D_{aa}(x-y)$ has a pole at
$q=(\omega,{\bf q})=0$ in the limit $\lambda\rightarrow 0$.
We set $M=T_m\in\{T_a\}$ and assume that
$\Delta_m\equiv \langle\psi^{\dag}T_m\psi\rangle\neq 0$
in the limit $\lambda\rightarrow 0$.
Then, we can show
\begin{equation}
\Delta_m= \frac{1}{\sum_cf_{amc}^2}\lambda
D_{aa}(\omega=0,{\bf q}=0),
\end{equation}
where we assumed that $[T_a,T_m]\neq 0$ and $f_{amc}$ are the structure 
constants defined by
\begin{equation}
[T_a,T_b]=\sum_cif_{abc}T_c.
\end{equation}
This indicates that
\begin{equation}
D_{aa}(\omega=0,{\bf q}=0)\propto 1/\lambda.
\end{equation}
Hence, we have a gapless mode in the limit $\lambda\rightarrow 0$.

The Higgs boson means the fluctuation mode of the amplitude of the
order parameter. 
We define the Higgs boson as
\begin{equation}
h = \psi^{\dag}T_m\psi.
\end{equation}

We show examples of spontaneous symmetry breaking in Table I.
Our formulation can be applied to second-order phase transitions
that occur as a spontaneous symmetry breaking.
An application to superconductivity is shown in the next section.
In the ferromagnetic transition, $SU(2)$ symmetry is broken to
$U(1)$ symmetry, where the bases of Lie algebra $\{T_a\}$ are given
by Pauli matrices $\sigma_a$.  For the Hubbard model\cite{hub63,mor85,yam98,yan96,yan16c},
the Lagrangian including the interaction term 
$V=U\psi_{\uparrow}^{\dag}\psi_{\uparrow}\psi_{\downarrow}^{\dag}
\psi_{\downarrow}$ is invariant under the transformation
$\psi\rightarrow e^{-i\theta\sigma_a}\psi$ for
$\psi= ^t(\psi_{\uparrow},\psi_{\downarrow})$ and $a=1$,2 and 3.
The symmetry breaking term is given as the magnetization of
electrons: $\mathcal{L}_{SB}=\lambda\psi^{\dag}\sigma_3\psi$.

\begin{table}
\caption{Examples of spontaneous symmetry breaking.
BEC indicates the Bose-Einstein condensation.
NJL denotes the Nambu-Jona-Lasinio model.
For superconductivity $\psi$ indicates the Nambu representation
$\psi(x)= ^t(\psi_{\uparrow}(x),\psi^{\dag}_{\downarrow})$.
For ferromagnetism $\psi$ is a doublet of fermions given by
$\psi(x)= ^t(\psi_{\uparrow}(x), \psi_{\downarrow}(x))$.
For BEC,$\phi$ is a complex scalar field.
In NJL $\psi$ is a Dirac spinor where 
$\bar{\psi}=\psi^{\dag}\gamma_0$.
}
\begin{center}
\begin{tabular}{|c|c|c|c|}
\hline
Phenomena & Symmetry & Higgs boson & NG boson  \\
\hline
Superconductivity & $\psi\rightarrow\exp(i\theta\sigma_3)\psi$ &
$\psi^{\dag}\sigma_1\psi$ & $\psi^{\dag}\sigma_2\psi$ \\
Ferromagnetism & $\psi\rightarrow\exp(i\theta\sigma_1)\psi$  &
$\psi^{\dag}\sigma_3\psi$ & $\psi^{\dag}\sigma_2\psi$ \\
BEC  &  $\phi\rightarrow \exp(i\theta)\phi$ &  $\phi+\phi^{\dag}$ &
$i(\phi-\phi^{\dag})$    \\
NJL  &  $\psi\rightarrow\exp(i\theta\gamma_5)\psi$ & $\bar{\psi}\psi$ &
$i\bar{\psi}\gamma_5\psi$ \\
\hline
\end{tabular}
\end{center}
\end{table}

\section{Nambu-Goldstone Green's function in superconductors}

\subsection{Hamiltonian and excitation modes}

Let us investigate Nambu-Goldstone modes in single-band
as well as multi-band superconductors.
This subject has been studied 
intensively\cite{suh59,per62,kon63,sta10,tan10a,tan10b,dia11,yan12,hu12,sta12,pla12,mai13,wil13,gan14,yer15,yan13,lin12,kob13,koy14,yan14,tan15,leg66}.
The Hamiltonian is
\begin{eqnarray}
H&=& \sum_{i\sigma}\int d{\bf r}\psi_{n\sigma}^{\dag}({\bf r})K_n({\bf r})
\psi_{n\sigma}({\bf r})\nonumber\\
&&- \sum_{ij}g_{nm}\int d{\bf r}\psi_{n\uparrow}^{\dag}({\bf r})
\psi_{n\downarrow}^{\dag}({\bf r})\psi_{m\downarrow}({\bf r})
\psi_{m\uparrow}({\bf r}),
\end{eqnarray}
where $n$ and $m$ (=1,2,$\cdots,N$)  are band indices.
$K_n({\bf r})$ stands for the kinetic operator given by
$K_n({\bf r})= p^2/(2m_n)-\mu\equiv\xi_n({\bf p})$ where $\mu$ is the
chemical potential.
We assume that $g_{nm}=g_{mn}^*$.
The second term indicates the pairing interaction with the coupling
constants $g_{nm}$.
This model is a simplified version of multi-band model and
the coupling constants $g_{nm}$
are assumed to be real constants.

We use the Nambu representation
\begin{eqnarray}
\psi_n= \left(
\begin{array}{c}
\psi_{n\uparrow} \\
\psi^{\dag}_{n\downarrow} \\
\end{array}
\right).
\end{eqnarray}
In the sigle-band case, the Hamiltonian is invariant under
the gauge transformation
\begin{equation}
\psi\rightarrow e^{-i\theta\sigma_3}\psi,
\end{equation}
for $\psi=\psi_1$.
This means the model has $U(1)$ phase invariance for
$\psi_{\sigma}\rightarrow e^{-i\theta}\psi_{\sigma}$.
In the multi-band case, however, the inter-band interactions
for $n\neq m$ break the gauge invariance.
Thus we have only one gauge invariance.  This indicates the existence of
one massless NG mode and other NG fields become massive modes.

In the $N$-band model, we have $N$ order parameters $\Delta_n$
$(n=1,\cdots,N)$.  The symmetry breaking terms are given by
\begin{equation}
H_{SB} = \sum_n \lambda_n\psi^{\dag}_n\sigma_1\psi_n,
\end{equation}
where  $\psi_n$ is the Nambu representation for the $n$-th band
fermions: $\psi_n = ^{t}(\psi_{n\uparrow},\psi^{\dag}_{n\downarrow})$.
$\lambda_n$ is an infinitesimal parameter for the $n$-th band.
According to the discussion in the previous section, the Nambu-Goldstone 
fields are
\begin{equation}
\pi_n = \psi_n^{\dag}\sigma_2\psi_n.
\end{equation}
The Higgs modes are represented by
\begin{equation}
h_n = \psi_n^{\dag}\sigma_1\psi_n,
\end{equation}
for $n=1,\dots,N$.
$h_n$ indicates the fluctuation mode of the amplitude of the
order parameter.

\subsection{Green's functions of Nambu-Goldstone modes}

\begin{figure}[htbp]
\begin{center}
\includegraphics[height=1.3cm]{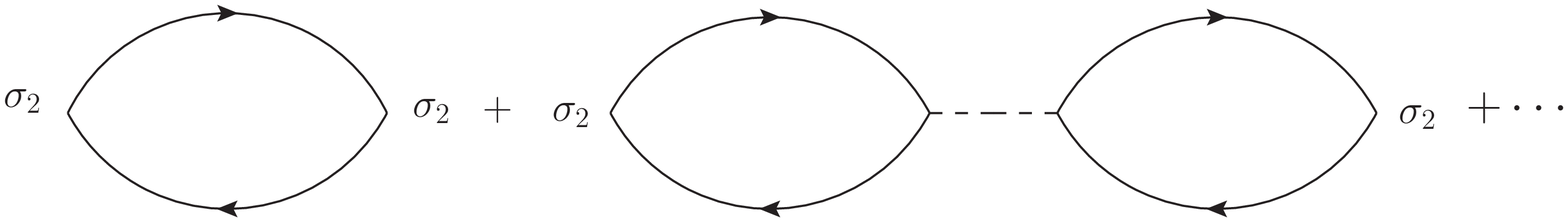}
\caption{Diagrams for Nambu-Goldstone boson Green's function.
The solid line shows the electron propagator and the dashed line denotes
the BCS pairing interaction.
}
\label{NGG}
\end{center}
\end{figure}

\begin{figure}[htbp]
\begin{center}
\includegraphics[height=1.8cm]{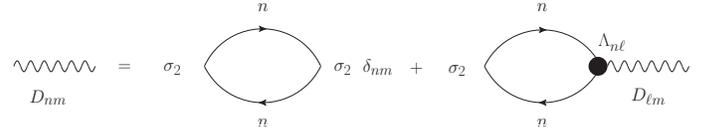}
\caption{Diagrams for the equation of  Nambu-Goldstone boson Green's 
function.
The solid line shows the electron propagator and the wavy line denotes
the NG Green function.  We introduced the vertex function indicated
by the dot.
}
\label{NGEq}
\end{center}
\end{figure}

The NG boson Green's functions are given as a matrix
$D=(D_{nm})$ where
\begin{equation}
D_{nm}(x-y) = -i\langle T\pi_n(x)\pi_m(y)\rangle.
\end{equation}
The Fourier transform of $D_{nm}(x-y)$ is denoted as $D_{nm}(\omega,{\bf q})$.
We show diagrams that contribute to NG boson Green's function in Fig.\ref{NGG}.
The equation for NG boson Green's functions is written as shown in 
Fig.\ref{NGEq}\cite{koy16,yan17}:
\begin{eqnarray}
D_{nm}(q)&=& -i\int\frac{dk_0}{2\pi}\frac{d^dk}{(2\pi)^d}\mathrm{tr}\Big[
4\delta_{nm}\sigma_2 G_n(k)\sigma_2 G_n(k+q)
\nonumber\\
&& -2\sum_{\ell}\sigma_2G_n(k)\Lambda_{n\ell}(k,k+q)
G_n(k+q)D_{\ell m}(q)\Big],\nonumber\\
\end{eqnarray}
for $q=(\omega,{\bf q})$
where $d$ is the space dimension, $\mathrm{tr}$ means taking
the trace of a matrix and $\Lambda_{n\ell}$ indicates 
the vertex function. 
The electron Green's function is defined as
\begin{eqnarray}
G_n(\omega,{\bf k})^{-1}= \left(
\begin{array}{cc}
\omega-\xi({\bf k}) & -\bar{\Delta}_n \\
-\bar{\Delta}_n & \omega+\xi({\bf k}) \\
\end{array}
\right)-\Sigma_n(k),
\end{eqnarray}
where the gap function $\bar{\Delta}_n$ is adopted to be real.
We use the four momentum notation $k=(k_0,{\bf k})=(\omega,{\bf k})$.
$\Sigma_n(k)$ indicates the self-energy part in the $n$-th band, 
given by a $2\times 2$
matrix, that represents an interaction effect.
In the finite-temperature formulation using the Matsubara
Green's functions, the integral with respect to $k_0$ is
replaced by the Matsubara frequency summation.

There is a relation between the vertex function $\Lambda_{n\ell}$
and the self-energy $\Sigma_n$ based on the Ward-Takahashi
identity\cite{koy16,yan17}.  We put the vertex function 
in the form: 
\begin{equation}
\Lambda_{n\ell}= \frac{1}{4}\tilde{\Lambda}_{n}g_{n\ell}\sigma_2,
\end{equation}
where $\tilde{\Lambda}_n$ shows a modification due to the self-energy
correction.  For the BCS model without electron interaction,
we have $\tilde{\Lambda}_n=1$.
We define
\begin{equation}
\chi_{NG,n}(q)= -\frac{1}{2}i\int\frac{dk_0}{2\pi}
\frac{d^dk}{(2\pi)^d}\mathrm{tr} \sigma_2G_n(k)\sigma_2 G_n(k+q),
\end{equation}
\begin{eqnarray}
\tilde{\chi}_{NG,n}(q)&=& -\frac{1}{2}i\int\frac{dk_0}{2\pi}
\frac{d^dk}{(2\pi)^d}\mathrm{tr} \tilde{\Lambda}_n(k,p)\sigma_2 G_n(k)
\sigma_2 \nonumber\\
&& \times  G_n(k+q). 
\end{eqnarray} 
We use the matrix notation $D=(D_{nm})$ for the NG boson Green's
function, and those for $\chi_{NG,n}$ and $\tilde{\chi}_{NG,n}$:
\begin{eqnarray}
\chi&=& {\rm diag}(\chi_{NG,1},\chi_{NG,2},\cdots),\\
\tilde{\chi}&=&{\rm diag}(\tilde{\chi}_{NG,1},\tilde{\chi}_{NG,2},\cdots).
\end{eqnarray}
The NG boson Green function matrix is represented as
\begin{equation}
D(q)= 8g^{-1}(g^{-1}+\tilde{\chi}(q))^{-1}\chi(q),
\end{equation}
where $g$ is the matrix of coupling constants given by $g=(g_{n\ell})$.
For a single-band superconductor, $g$ itself denotes the
coupling constant.
The singularity of $D$ is determined by the zero of
$g^{-1}+\tilde{\chi}$.

We can show that a massless mode indeed exists, namely, the
dispersion $\omega({\bf q})$ approaches zero as 
$|{\bf q}|\rightarrow 0$.
We consider the case where the self-energy part $\Sigma_n$ vanishes.
$\chi(q)$ can be calculated exactly in the limit ${\bf q}=0$.
The inverse of $D(q)$ for $q=(\omega,{\bf q})$ is proportional to
\begin{eqnarray}
&& (D^{-1}(\omega,{\bf q}))_{nm} \nonumber\\ 
 &\propto& (g^{-1})_{nm}-\frac{\delta_{nm}}{2}
i{\rm Tr}\sigma_2G_n^{(0)}(k)\sigma_2
 G_n^{(0)}(k+q)\nonumber\\
&=& (g^{-1})_{nm}-\delta_{nm}\Big[ f_n
+\frac{1}{2}N_n(0)\frac{\omega}{\bar{\Delta}_n}
 \frac{1}{\sqrt{1-(\omega/2\bar{\Delta}_n)^2}} \nonumber\\
&& \times {\rm tan}^{-1}\sqrt{\frac{(\omega/2\bar{\Delta}_n)^2}{1-(\omega/2\bar{\Delta}_n)^2}}
+ O(|{\bf q}|^2) +\cdots \big],
\end{eqnarray}
for $|{\bf q}|\rightarrow 0$
where $\bar{\Delta}_n$ is the mean-field value of the order parameter  
and $\delta_{nm}$ is the Kronecker delta.  $v_{Fn}$ is the
Fermi velocity of the $n$-th band, and $N_n(0)$ is the density of
states at the Fermi surface in the $n$-th band.
We used the notation
\begin{equation}
f_n = N_n(0)\int d\xi_k\frac{1}{2E_n}{\rm tanh}\left(\frac{E_n}{2k_BT}\right),
\end{equation}
for $E_n=\sqrt{\xi_k^2+\bar{\Delta}_n^2}$.
Because the gap equation is given as
\begin{equation}
{\rm det}( g^{-1}-F)=0,
\end{equation}
where $F={\rm diag}(f_1,\cdots,f_N)$ is the diagonal matrix with elements $f_n$
($n=1,\cdots,N$),
${\rm det}D^{-1}(\omega,{\bf q})$ has a zero as $\omega\rightarrow 0$
and ${\bf q}\rightarrow 0$, indicating that
$\omega({\bf q})\rightarrow 0$ as ${\bf q}\rightarrow 0$.  
Thus the NG mode exists with vanishing gap.
The other $N-1$ modes become massive due to the interband
couplings $g_{nm}$.
These modes are called the Leggett modes\cite{leg66}.
When all the couplings $g_{nm}$ vanish, we have $N$ massless modes.

\subsection{Poles of the NG boson Green's function $D$}

The energy dispersion of the NG mode is determined by the poles
of the NG Green function. 
In general, we have one massless mode and $N-1$ gapped modes
for an $N$-band superconductor.
The poles depend on the intraband and interband couplings $g_{nm}$.
We consider the case $\Sigma_n=0$ in this subsection.
In this case we have $\tilde{\chi}=\chi$.

We assume that $v_{Fn}|{\bf q}|/q_0$ is small for $q=(q_0,{\bf q})$.
At absolute zero, we obtain
\begin{eqnarray}
\chi_{NG,n}(q)
&=& -N_n(0)\int d\xi_k\frac{1}{2E_n(\xi_k)} \nonumber\\
&& -N_n(0)\frac{\tilde{q}_n}{2\bar{\Delta}_n}
\frac{1}{ \sqrt{1-(\tilde{q}_n/2\bar{\Delta}_n)^2} }\nonumber\\
&& \times \tan^{-1}\left( \frac{\tilde{q}_n}{2\bar{\Delta}_n}
\frac{1}{ \sqrt{1-(\tilde{q}_n/2\bar{\Delta}_n)^2}}\right),
\end{eqnarray}
for 
\begin{equation}
\tilde{q}_n= \sqrt{q_0^2-\frac{1}{3}v_{Fn}^2{\bf q}^2 },
\end{equation}
where we used an approximation 
$({\bf v}_F\cdot{\bf q})^2\approx (1/3)v_F^2|{\bf q}|^2$.
$q_0$ and ${\bf q}$ appear as a linear combination
$q_0^2-(1/3)v_{Fn}^2{\bf q}^2$ for $v_{Fn}|{\bf q}|/q_0\ll 1$.
The equation ${\rm det}(g^{-1}+\tilde{\chi})(q)=0$ determines the
dispersion relation of the NG modes.
We have one massless mode since ${\rm det}D^{-1}(q)$ vanishes
as $q=(q_0,{\bf q})\rightarrow 0$.
$D(q)$ has a pole with a finite residue.
Using $\tan^{-1}x=x-(1/3)x^3+\cdots$,
for a single-band superconductor, the NG Green function for small $q_0$ 
and ${\bf q}$ is written as
\begin{equation}
D(q)\simeq 8N(0)\frac{1}{(gN(0))^2}
\frac{(2\Delta)^2}{q_0^2-c_s^2{\bf q}^2},
\end{equation}
where $c_s=v_F/\sqrt{3}$.
In the two-band case, we obtain
\begin{eqnarray}
D(q)&=& \tilde{D}(q)\Big[ \frac{\gamma_{12}}{4\bar{\Delta}_1\bar{\Delta}_2}(N_1+N_2)
(q_0^2-v_{NG}^2{\bf q}^2)\nonumber\\
&&  +\frac{N_1N_2}{16\bar{\Delta}_1^2\bar{\Delta}_2^2}(q_0^4
-\frac{1}{3}(v_{F1}^2+v_{F2}^2)q_0^2{\bf q}^2) \Big]^{-1},
\end{eqnarray}
where
\begin{equation}
v_{NG}^2 = \frac{1}{3}
\frac{N_1v_{F1}^2+N_2v_{F2}^2}{N_1+N_2},
\end{equation}
and we write $N_m=N_m(0)$.
$\gamma_{nm}$ is the strength of Josephson coupling defined as the inverse
of $g$:
\begin{equation}
\gamma_{nm}=(g^{-1})_{nm}.
\end{equation}
There are corrections to the coefficient of $q_0^4$ and $q_0^2{\bf q}^2$, 
which are of the order of $\gamma_{12}$.  We neglected them in eq.(36).
$\tilde{D}(q)$ is a matrix given as
\begin{eqnarray}
\tilde{D}(q)&=& g^{-1}\left(
\begin{array}{cc}
\gamma_{22}-f_2-N_2I(\frac{\tilde{q}_2}{2\bar{\Delta}_2})  &  -\gamma_{12} \\
-\gamma_{21}  &  \gamma_{11}-f_1-N_1I(\frac{\tilde{q}_1}{2\bar{\Delta}_1}) \\
\end{array}
\right) \nonumber\\
&&\times \left(
\begin{array}{cc}
\chi_1(q) & 0 \\
0 & \chi_2(q) \\
\end{array}
\right),\nonumber\\
\end{eqnarray}
where we put
\begin{equation}
I(x)= \frac{x}{\sqrt{1-x^2}}\tan^{-1}\left(\frac{x}{\sqrt{1-x^2}}\right).
\end{equation}
In deriving $D(q)$ in eq.(36), we used the gap equation written as
\begin{eqnarray}
(\gamma_{11}-f_1)\bar{\Delta}_1+\gamma_{12}\bar{\Delta}_2 &=& 0,\\
\gamma_{21}\bar{\Delta}_1+(\gamma_{22}-f_2)\bar{\Delta}_2 &=& 0.
\end{eqnarray}
From zeros of the denominator of eq.(36), the dispersions of
excitation modes (Nambu-Goldstone-Leggett modes) are determined.
The Nambu-Goldstone mode has the dispersion
$\omega_{NG}({\bf q})= v_{NG}|{\bf q}|$,
and the massive mode has\cite{sha02}
\begin{equation}
\omega_{NGL}({\bf q})^2= \omega_L^2+v_L^2{\bf q}^2,
\end{equation}
where
\begin{equation}
\omega_L = \sqrt{ \frac{4|\gamma_{12}\bar{\Delta}_1\bar{\Delta}_2|\frac{N_1+N_2}{N_1N_2}}
{1+\frac{2}{3}\gamma_{12}\left(\frac{1}{N_1}\frac{\bar{\Delta}_1}{\bar{\Delta}_2}+
\frac{1}{N_2}\frac{\bar{\Delta}_2}{\bar{\Delta}_2}\right)}},
\end{equation}
\begin{eqnarray}
v_L^2 &=& \frac{1}{3}\frac{N_1v_{F2}^2+N_2v_{F1}^2}{N_1+N_2}\nonumber\\
&+&\frac{2}{9}v_{F1}^2\left( \frac{\gamma_{12}}{N_2}\frac{\bar{\Delta}_2}{\bar{\Delta}_1}
-\frac{\gamma_{12}}{N_1}\frac{\bar{\Delta}_1}{\bar{\Delta}_2} \right)
\nonumber\\
&& \times
\left( 1+\frac{2}{3}\gamma_{12}\left(\frac{1}{N_1}\frac{\bar{\Delta}_1}{\bar{\Delta}_2}+
\frac{1}{N_2}\frac{\bar{\Delta}_2}{\bar{\Delta}_2}\right)\right)^{-1},
\nonumber\\
&+&\frac{2}{9}v_{F2}^2\left( \frac{\gamma_{12}}{N_1}\frac{\bar{\Delta}_1}{\bar{\Delta}_2}
-\frac{\gamma_{12}}{N_2}\frac{\bar{\Delta}_2}{\bar{\Delta}_1} \right)
\nonumber\\
&& \times
\left( 1+\frac{2}{3}\gamma_{12}\left(\frac{1}{N_1}\frac{\bar{\Delta}_1}{\bar{\Delta}_2}+
\frac{1}{N_2}\frac{\bar{\Delta}_2}{\bar{\Delta}_2}\right)\right)^{-1}.
\nonumber\\
\end{eqnarray}
We included a correction of the order of $\gamma_{12}$.
Here please note that
when $\gamma_{12}<0$, we have $\bar{\Delta}_1\bar{\Delta}_2>0$, and
when $\gamma_{12}>0$, $\bar{\Delta}_1\bar{\Delta}_2<0$ from the gap equation.
Thus, $\gamma_{12}\bar{\Delta}_1\bar{\Delta}_2<0$.

We show the excitation energy as a function of $g_{12}$
for two-band superconductors in Fig. 3.
There are two massless modes when $g_{12}=0$, and the one mode
becomes massive for non-zero $g_{12}$.
The excitation energy $\omega_L$ of the Leggett mode is dependent upon
coupling constant $\{g_{nm}\}$.
We show $\omega_L$ as a function of $g_{22}$ in Fig. 4 for
a two-band superconductor.
The excitation gap energy is determined as a zero of
${\rm det}D^{-1}(\omega,{\bf q}=0)$ which is shown in Fig. 5.

For a three-band superconductor, ${\rm det}(g^{-1}+\chi)$
is expanded as follows for small $q_0$ and ${\bf q}$:
\begin{eqnarray}
&& {\rm det}(g^{-1}+\chi) \nonumber\\
&=& -\frac{N_1+N_2+N_3}{4(\bar{\Delta}_1\bar{\Delta}_2\bar{\Delta}_3)^2}
(\gamma_{12}\gamma_{13}\bar{\Delta}_1^2\bar{\Delta}_2\bar{\Delta}_3
 +\gamma_{12}\gamma_{23}\bar{\Delta}_1\bar{\Delta}_2^2\bar{\Delta}_3 
\nonumber\\
&+& \gamma_{13}\gamma_{23}\bar{\Delta}_1\bar{\Delta}_2\bar{\Delta}_3^2)
(q_0^2-v_{NG}^2{\bf q}^2) \nonumber\\
&& -\frac{1}{16(\bar{\Delta}_1\bar{\Delta}_2\bar{\Delta}_3)^2}
\big[ N_2(N_1+N_3)\gamma_{13}\bar{\Delta}_1\bar{\Delta}_3 \nonumber\\
&& +N_1(N_2+N_3)\gamma_{23}\bar{\Delta}_2\bar{\Delta}_3 
+ N_3(N_1+N_2)\gamma_{12}\bar{\Delta}_1\bar{\Delta}_2 \big]q_0^4\nonumber\\
&-& \frac{1}{64(\bar{\Delta}_1\bar{\Delta}_2\bar{\Delta}_3)^2}
N_1N_2N_3 q_0^6 \nonumber\\
&+& \frac{1}{16(\bar{\Delta}_1\bar{\Delta}_2\bar{\Delta}_3)^2}
\big[ N_1N_2(\gamma_{13}\bar{\Delta}_1\bar{\Delta}_3
+\gamma_{23}\bar{\Delta}_2\bar{\Delta}_3)\nonumber\\
&& \times\frac{1}{3}(v_{F1}^2+v_{F2}^2)
+N_2N_3(\gamma_{12}\bar{\Delta}_1\bar{\Delta}_2
+\gamma_{13}\bar{\Delta}_1\bar{\Delta}_3)\nonumber\\
&& \times\frac{1}{3}(v_{F2}^2+v_{F3}^2)
+N_1N_3(\gamma_{12}\bar{\Delta}_1\bar{\Delta}_2
+\gamma_{23}\bar{\Delta}_2\bar{\Delta}_3)\nonumber\\
&& \times\frac{1}{3}(v_{F1}^2+v_{F3}^2) \big]q_0^2{\bf q}^2
+\cdots,
\end{eqnarray} 
where we assume $\gamma_{nm}=\gamma_{mn}$ and used the gap equation
for three gaps:
\begin{eqnarray}
\left(
\begin{array}{ccc}
\gamma_{11}-f_1  &  \gamma_{12}  &  \gamma_{13} \\
\gamma_{21}  &  \gamma_{22}-f_2  & \gamma_{23}  \\
\gamma_{31}  & \gamma_{32}  & \gamma_{33}-f_3  \\
\end{array}
\right)\left(
\begin{array}{c}
\bar{\Delta}_1 \\
\bar{\Delta}_2 \\
\bar{\Delta}_3 \\
\end{array}
\right)=0.
\end{eqnarray}
The velocity for the NG mode in the three-band case is given by
\begin{equation}
v_{NG}^2= \frac{1}{3}\frac{N_1v_{F1}^2+N_2v_{F2}^2+N_3v_{F3}^2}{N_1+N_2+N_3}.
\end{equation}
The velocity
$v_{NG}$ in the $N$-band case is straightforwardly obtained as
\begin{equation}
v_{NG}^2= \frac{1}{3}\frac{N_1v_{F1}^2+\cdots+N_Nv_{FN}^2}{N_1+\cdots+N_N}.
\end{equation}
The gap $\omega_L$ of the massive mode is obtained as a zero of
${\rm det}(g^{-1}+\chi)$ in the limit ${\bf q}\rightarrow 0$, that is,
a solution of the cubic equation  for $q_0^2$.  $\omega_L$ is given as
\begin{equation}
\omega_L^2= -\frac{2A_1}{A_0}\pm \frac{2}{A_0}\sqrt{A_1^2-4A_0A_2},
\end{equation}
where
\begin{eqnarray}
A_0 &=& N_1N_2N_3,\\
A_1 &=& N_1(N_2+N_3)\gamma_{23}\bar{\Delta}_2\bar{\Delta}_3
+ N_2(N_1+N_3)\gamma_{13}\bar{\Delta}_1\bar{\Delta}_3  \nonumber\\
&& + N_3(N_1+N_2)\gamma_{12}\bar{\Delta}_1\bar{\Delta}_2, \\ 
A_2 &=& (N_1+N_2+N_3)\bar{\Delta}_1\bar{\Delta}_2\bar{\Delta}_3
(\gamma_{12}\gamma_{13}\bar{\Delta}_1+\gamma_{12}\gamma_{23}\bar{\Delta}_2
\nonumber\\
&& +\gamma_{13}\gamma_{23}\bar{\Delta}_3).
\end{eqnarray}
Here we assumed that $A_1<0$ and $A_1^2-4A_0A_2\ge 0$.
The lower energy gap is given by $(2/A_0)(|A_1|-\sqrt{A_1^2-A_0A_2})$.
If $A_1^2\gg A_0A_2$ is satisfied, this leads to
$\omega_L^2\approx 4A_2/|A_1|$.  When the contribution of one band is
small, say $N_3\ll N_1, N_2$, this gap reduces to that of the
Leggett mode in the two-band case.

In the three-band case, there are in general three zeros in
the determinant ${\rm det}D^{-1}(\omega,{\bf q}=0)$ as shown
in Fig. 6.  When $N\geq 3$, there is the case where several
Leggett modes become massless.  This is shown in Fig. 7
where the excitation energy $\omega_L$ is presented as a function of
$g_{12}$.  $\omega_L$ vanishes and one mode becomes massless
when the time-reversal symmetry is broken.

Let us turn to examine $D(q)$ when $\tilde{q}_n\sim 2\bar{\Delta}_n$
($\tilde{q}_n\le 2\bar{\Delta}_n$).
An analytic property of $\chi(q)$ will be changed as $q_0$ approaches
the threshold energy $2\bar{\Delta}_n$.
Using $\tan^{-1}(1/t)\simeq \frac{\pi}{2}-t+\frac{1}{3}t^3+\cdots$
for small $t>0$,
we have
\begin{equation}
\chi_n(q)\simeq -f_n-N_n(0)\frac{\pi}{2}
\frac{2\bar{\Delta}_n}{ \sqrt{(2\bar{\Delta}_n)^2+\frac{1}{3}v_{Fn}^2{\bf q}^2
-q_0^2} },
\end{equation}
for $\tilde{q}_n=\sqrt{q_0^2-\frac{1}{3}v_{Fn}^2{\bf q}^2}\simeq 2\bar{\Delta}_n$.
In the single-band case, the Green function $D(q)$ is
\begin{equation}
D(q)\simeq \frac{8}{g}\left( 1+\frac{1}{gN(0)\pi\bar{\Delta}}
\sqrt{(2\bar{\Delta})^2+\frac{1}{3}v_F^2{\bf q}^2-q_0^2}+\cdots \right).
\end{equation}
At the threshold $\tilde{q}=2\bar{\Delta}$, $D(q)$ becomes a constant:
$D(q)\simeq 8/g$.
$D(q)$ has no pole as a function of $q_0$ and loses a quasiparticle
character in this region.

\begin{figure}[htbp]
\begin{center}
\includegraphics[width=7.5cm]{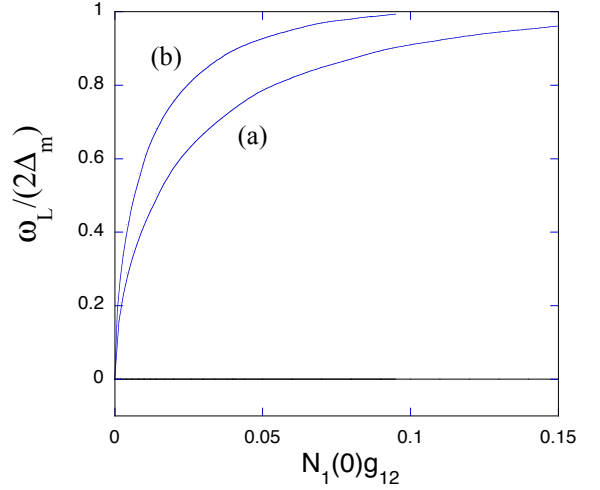}
\caption{Gap energy of the Nambu-Goldstone-Leggett mode
as a function of $N_1(0)g_{12}$ for two-band superconductors.
The parameters are (a) $N_1(0)g_{11}=N_2(0)g_{22}=0.3$, and
(b) $N_1(0)g_{11}=0.3$ and $N_2(0)g_{22}=0.25$.
We set $\Delta_m=min( \bar{\Delta}_1,\bar{\Delta}_2)$ and
$N_1(0)=N_2(0)$.
}
\label{NG1}
\end{center}
\end{figure}

\begin{figure}[htbp]
\begin{center}
\includegraphics[width=7.5cm]{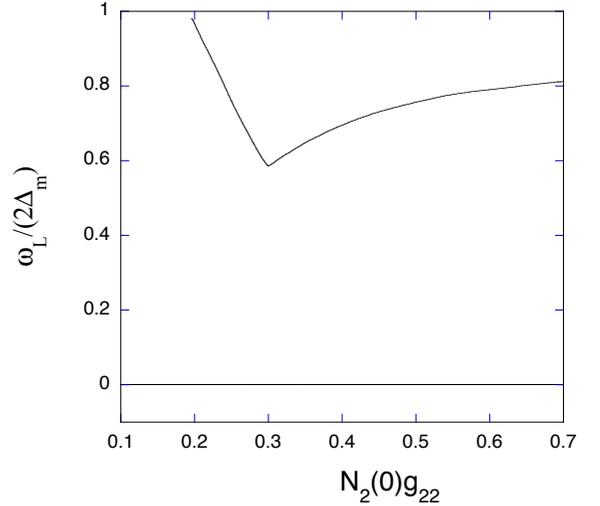}
\caption{Gap energy of the Nambu-Goldstone-Leggett mode
as a function of $N_2g_{22}$ for a two-band superconductor.
We used $N_1g_{11}=0.3$ and $N_1g_{12}=0.02$.
We set $\Delta_m=min( \bar{\Delta}_1,\bar{\Delta}_2 )$ and
$N_1(0)=N_2(0)$.
}
\label{NG2}
\end{center}
\end{figure}

\begin{figure}[htbp]
\begin{center}
\includegraphics[width=7.5cm]{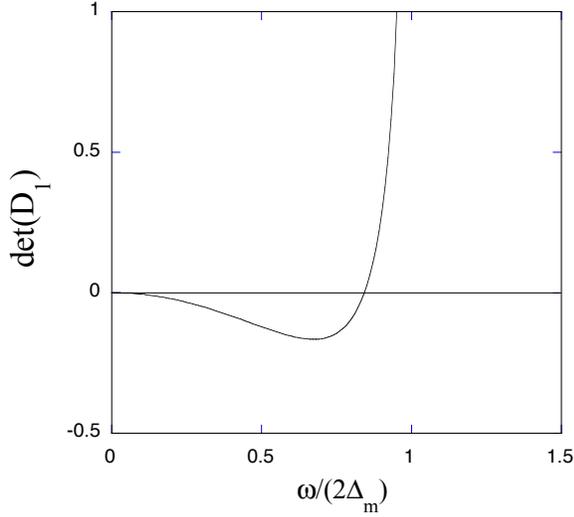}
\caption{The determinant ${\rm det}D_1(\omega,{\bf q}=0)$ as a function of
$\omega/(2\Delta_m)$ where $D_1=g^{-1}+\chi$.
The zero of ${\rm det}D_1$ indicates the excitation energy.
We used $N_1g_{11}=0.3$, $N_2g_{22}=0.25$ and $N_1g_{12}=0.03$.
We set $\Delta_m=min( \bar{\Delta}_1,\bar{\Delta}_2 )$ and
$N_1(0)=N_2(0)$.
}
\label{NG3}
\end{center}
\end{figure}

\begin{figure}[htbp]
\begin{center}
\includegraphics[width=7.5cm]{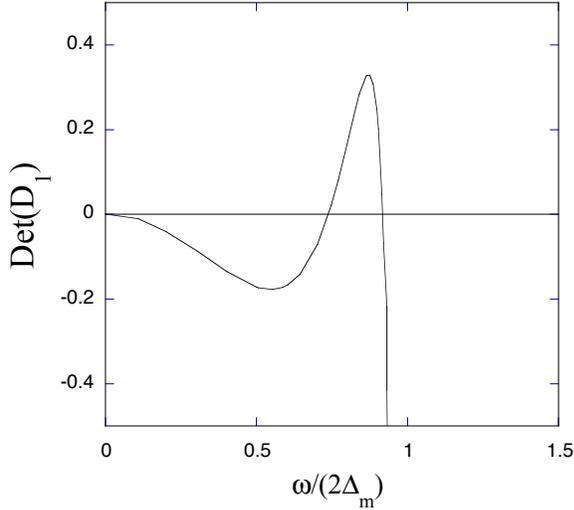}
\caption{The determinant ${\rm det}D_1(\omega,{\bf q}=0)$ as a function of
$\omega/(2\Delta_m)$ for $D_1=g^{-1}+\chi$ in a three-gap case.
The zero of ${\rm det}D_1$ indicates the excitation energy,
There are three zeros in the three-gap case.
We used $N_1g_{11}=N_2g_{22}=N_3g_{33}=0.3$, $N_1g_{12}=0.05$,
$N_1g_{23}=0.04$ and $N_1g_{13}=0.02$.
We set $\Delta_m=min( \bar{\Delta}_1,\bar{\Delta}_2,\bar{\Delta}_3 )$ and
$N_1(0)=N_2(0)=N_3(0)$.
}
\label{NG4}
\end{center}
\end{figure}

\begin{figure}[htbp]
\begin{center}
\includegraphics[width=7.5cm]{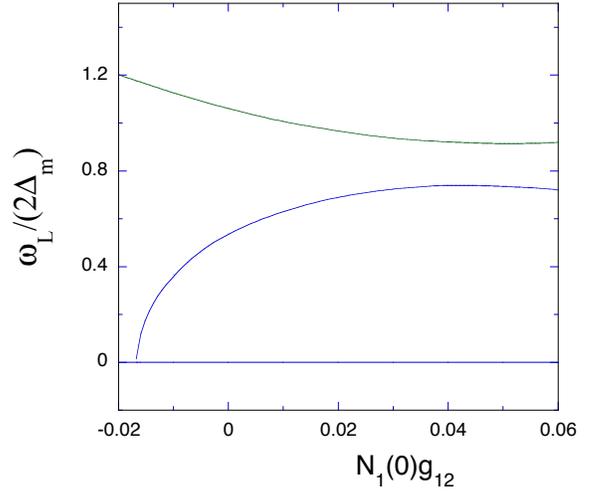}
\caption{Gap energy of the Nambu-Goldstone-Leggett mode
as a function of $N_1g_{12}$ in a three-gap case.
The zero of ${\rm det}D_1$ indicates the excitation energy,
There are three zeros in the three-gap case.
We used $N_1g_{11}=N_2g_{22}=N_3g_{33}=0.3$, $N_1g_{23}=0.05$,
and $N_1g_{23}=0.02$.
We set $\Delta_m=min( \bar{\Delta}_1,\bar{\Delta}_2,\bar{\Delta}_3 )$ and
$N_1(0)=N_2(0)=N_3(0)$.
}
\label{NG5}
\end{center}
\end{figure}

\subsection{Ward-Takahashi identity and impurity scattering}

We investigate the effect of impurity scattering on the Nambu-Goldstone
modes in superconductors.  
The Hamiltonian for impurity scattering that we consider is written as
\begin{eqnarray}
H_{imp}&=& \sum_i\sum_{m\sigma}u_{1mm}({\bf r}-{\bf R}_i)
\psi_{m\sigma}^{\dag}({\bf r})\psi_{m\sigma}({\bf r}) \nonumber\\
&+& \sum_i\sum_m u_{2m}({\bf r}-{\bf R}_i){\bf S}_i\cdot {\bf s}_{mi}
\nonumber\\
&\equiv& \sum_{m}\sum_{\sigma\sigma'}\psi_{m\sigma}^{\dag}
V_{m\sigma\sigma'}\psi_{m\sigma'}.
\end{eqnarray}
where ${\bf R}_i$ indicates the position of an impurity,
${\bf S}_i$ indicates the spin operator of the impurity spin and
${\bf s}_{mi}$ is the spin operator of the electron in the
$m$-th band.
Here we consider only the intra band scattering for simplicity
because the interband scattering breaks the gauge invariance
under 
$\psi_{m\sigma}\rightarrow e^{-i\theta_m}\psi_{m\sigma}$.

We introduce the matrix self-energy to
take account of the electron scattering due to non-magnetic and
magnetic impurities:
\begin{eqnarray}
\Sigma_m=\left(
\begin{array}{cc}
\Sigma_{1m}  & \Sigma_{2m} \\
\Sigma_{2m}^* & \Sigma_{1m} \\
\end{array}
\right).
\end{eqnarray}
We write the Green function in the form,
\begin{eqnarray}
G_m(k)= \left(
\begin{array}{cc}
G_{m\uparrow\uparrow}(k)  &  F_m(k) \\
F_m^*(k)  &  G_{m\downarrow\downarrow}(k) \\
\end{array}
\right).
\end{eqnarray}
Here we use the finite-temperature formalism by putting
$k_0\rightarrow i\omega_n$ where $\omega_n=(2n+1)\pi/\beta$
is the Matsubara frequency\cite{agd}.
In the impurity scattering, self-energies are given as\cite{kop01}
\begin{eqnarray}
\Sigma_{1m}(i\omega_n,{\bf k}) &=& \frac{1}{2\pi\tau_{1m}}\int d\xi_k 
G_{m\uparrow\uparrow}(i\omega_n,{\bf k}),\\
\Sigma_{2m}(i\omega_n,{\bf k}) &=& \frac{1}{2\pi\tau_{2m}}\int d\xi_k 
F_m(i\omega_n,{\bf k}),
\end{eqnarray}
where $\tau_{1m}$ and $\tau_{2m}$ indicate relaxation times
in the $m$-th band due to
impurity scattering.
We have $\tau_{1m}=\tau_{2m}$ for scattering by non-magnetic impurities,
and $\tau_{1m}\neq \tau_{2m}$ for magnetic impurities.

$\Sigma_{1m}$ and $\Sigma_{2m}$ are related as follows.
\begin{equation}
\Sigma_{1m}(i\omega_n) = -\frac{i}{2\tau_{1m}}
\frac{\omega_n+i\Sigma_{1m}}{\sqrt{(\omega_n^2+i\Sigma_{1m})^2
+(\bar{\Delta}_m+\Sigma_{2m})^2}}, \\ 
\end{equation}
\begin{equation}
\Sigma_{2m}(i\omega_n) = \frac{1}{2\tau_{2m}}
\frac{\bar{\Delta}_m+\Sigma_{2m}}{\sqrt{(\omega_n^2+i\Sigma_{1m})^2
+(\bar{\Delta}_m+\Sigma_{2m})^2}}.\\ 
\end{equation}
We define $\tau_{sm}$ by
\begin{equation}
\frac{1}{\tau_{sm}}= \frac{1}{\tau_{1m}}-\frac{1}{\tau_{2m}}.
\end{equation}
When $1/\tau_{sm}$ is small, $\Sigma_{1m}$ and $\Sigma_{2m}$
are expanded as
\begin{eqnarray}
-\Sigma_{1m}(i\omega_n) &=& \frac{1}{2\tau_{1m}}
\frac{i\omega_n}{\sqrt{\omega_n^2+\bar{\Delta}_m^2}} 
+ \frac{1}{4\tau_{1m}\tau_{sm}}
\frac{i\omega_n\bar{\Delta}_m^2}{(\omega_n^2+\bar{\Delta}_m^2)^2},
\nonumber\\
\end{eqnarray}
\begin{eqnarray}
\Sigma_{2m}(i\omega_n) &=& \frac{1}{2\tau_{2m}}
\frac{\Delta_n}{\sqrt{\omega_n^2+\bar{\Delta}_m^2}} 
- \frac{1}{4\tau_{2m}\tau_{sm}}
\frac{i\omega_n^2\bar{\Delta}_m}{(\omega_n^2+\bar{\Delta}_m^2)^2},
\nonumber\\
\end{eqnarray}
The inverse Green function is expressed as
\begin{eqnarray}
G_n^{-1}(i\omega,{\bf k})= \left(
\begin{array}{cc}
i\omega-\xi({\bf K})-\Sigma_{1m}  &  -\bar{\Delta}_n-\Sigma_{2n} \\
-\bar{\Delta}_n^*-\Sigma_{2n}^*  &  i\omega+\xi({\bf k})-\Sigma_{1n} \\
\end{array}
\right).\nonumber\\
\end{eqnarray}

The Ward-Takahashi identity is used to obtain a relation
between the self-energy and the vertex function.
The Ward-Takahashi identity is given as\cite{koy16,yan17}
\begin{eqnarray}
&& (q-k)_{\mu}G_n(k)\Gamma^{\mu}_{nm}(k,q)G_n(q) \nonumber\\
&& ~~~ = i\delta_{nm}\sigma_3G_n(q)-i\delta_{nm}(k)\sigma_3
-\lambda_mG_n(k)\Gamma_{nm}(k,q)G_n(q), \nonumber\\
\end{eqnarray}
where
\begin{equation}
\Gamma_{nm}(k,q)= -2\delta_{nm}\sigma_2
+\sum_{\ell}\Lambda_{n\ell}D_{\ell m}(q-k).
\end{equation}
When $q\rightarrow k$, we have
\begin{equation}
iG_n^{-1}(k)\sigma_3-i\sigma_3G_n^{-1}(k)-\sum_m
\lambda_m\Gamma_{nm}(k,k)=0.
\end{equation}
The relation in eq.(11) is generalized to the
multiband case as: 
\begin{equation}
\sum_m\lambda_mD_{nm}(q\rightarrow 0)= -8\bar{\Delta}_n,
\end{equation}
as $\lambda_m\rightarrow 0$.
Then we obtain
\begin{equation}
\sum_m\lambda_m\Gamma_{nm}(k,k)\rightarrow
-8\sum_{\ell}\Lambda_{n\ell}(k,k)\bar{\Delta}_{\ell}
\end{equation}
in the limit $\lambda_m\rightarrow 0$. 
This results in
\begin{equation}
iG_n^{-1}(k)\sigma_3-i\sigma_3G_n^{-1}(k)+8\sum_{\ell}
\Lambda_{n\ell}(k,k)\bar{\Delta}_{\ell}=0.
\end{equation}
The vertex function is determined as
$\Lambda_{n\ell}= (1/4)\tilde{\Lambda}_{n}g_{n\ell}\sigma_2$
with
\begin{equation}
\tilde{\Lambda}_{n}=1+\frac{\Sigma_{2n}}{\bar{\Delta}_n}
= 1+\frac{1}{2\tau_{2n}}\frac{1}{\sqrt{\omega^2+\bar{\Delta}_n^2}}.
\end{equation}
Here we used the gap equation given by
\begin{equation}
\bar{\Delta}_n= \sum_{\ell}g_{n\ell}\frac{1}{\beta}\sum_{\omega}
\int\frac{d^dk}{(2\pi)^d}F_{\ell}(i\omega,{\bf k}).
\end{equation}
From this relation, we can show that the NG boson Green function
$D(q)$ has a pole at $q=(q_0,{\bf q})=0$.
In fact, the matrix $g^{-1}+\tilde{\chi}(q)$ in the limit $q\rightarrow 0$
is represented as
\begin{eqnarray}
&&(g^{-1})_{n\ell}+\delta_{n\ell}\tilde{\chi}_{NG,n}(q=0) \nonumber\\
&=& (g^{-1})_{n\ell} \nonumber\\
 && -\delta_{n\ell}\frac{1}{\bar{\Delta}_n}N_n(0)\pi
\frac{1}{\beta}\sum_{\omega}
\frac{\bar{\Delta}_n+\Sigma_{2n}}
{\sqrt{ (\omega+i\Sigma_{1n})^2+(\bar{\Delta}_n+\Sigma_{2n})^2}}.
\nonumber\\
\end{eqnarray}
The determinant of this matrix vanishes because of the gap equation
written as
\begin{equation}
{\rm det}\left( (g^{-1})_{n\ell}-\delta_{n\ell}\frac{1}{\bar{\Delta}_n}
s_n\right)=0,
\end{equation}
where
\begin{equation}
s_n= \pi N_n(0)
\frac{1}{\beta}\sum_{\omega}\frac{\bar{\Delta}_n+\Sigma_{2n}}
{\sqrt{ (\omega+i\Sigma_{1n})^2+(\bar{\Delta}_n+\Sigma_{2n})^2}}.
\end{equation}

\section{Plasma mode}

The Nambu-Goldstone mode becomes a massive plasma mode in the
presence of the Coulomb potential.
The plasma mode is represented by the spatial derivative of
the phase variables $\theta_n$ where the order parameters are
parametrized as $\Delta_n= e^{i2\theta_n}(\bar{\Delta}_n+h_n)$.
The action density for $\theta_j$ is written as\cite{yan15,yan17}
\begin{eqnarray}
\mathcal{L}[\theta]&=& \sum_j \Big[ N_j(0)(\partial_{\tau}\theta_j-e\phi)^2
+n_j\frac{1}{2m_j}(\nabla\theta_j)^2 \Big] \nonumber\\ 
&+& \frac{1}{8\pi}(\nabla\phi)^2 
+ \sum_{ij} \bar{\Delta}_i(g^{-1})_{ij}\bar{\Delta}_j\cos(2(\theta_i-\theta_j)),
\nonumber\\
\end{eqnarray}
where $e$ is the charge of the electron and $\phi$ indicates the
Coulomb potential.
The effective action for $\theta_j$ is obtained by integrating out
the field $\phi$:
\begin{eqnarray}
\mathcal{L}[\theta]&=& \frac{1}{8\pi e^2N(0)^2}
\sum_{jj'a}N_j(0) N_{j'}(0)\partial_{\tau}\zeta_{ja}\partial_{\tau}\zeta_{j'a}
\nonumber\\
&&+ \sum_{ja}\frac{n_j}{2m_{j}}\zeta_{ja}^2
+ \sum_jN_j(0)\left(\partial_{\tau}\theta_j\right)^2
\nonumber\\
&& -\frac{1}{N(0)}\left( \sum_jN_j(0)\partial_{\tau}\theta_j \right)^2
\nonumber\\
&& + \sum_{ij} \bar{\Delta_i}(g^{-1})_{ij}\bar{\Delta_j}\cos(2(\theta_i-\theta_j))
+\cdots ,
\end{eqnarray}
where we put
\begin{equation}
\zeta_{ja}=\nabla_a\theta_j,
\end{equation}
and $N(0)=\sum_jN_j(0)$.  The index $a$
takes $x$, $y$ and $z$.

In the single-band case, the plasma frequency is given by
\begin{equation}
\omega_{pl,a}^2= 4\pi e^2 n/m_a,
\end{equation}
where $n=n_1$ is the electron density.
In the two-band case, the quadratic part of $\zeta_{ja}$ is written as
\begin{eqnarray}
\frac{1}{\beta}\sum_{\ell}\frac{1}{8\pi e^2N(0)^2}\sum_a
(\zeta_{1a}(\omega_{\ell})~\zeta_{2a}(\omega_{\ell}))
M_{\zeta}\left(
\begin{array}{c}
\zeta_{1a}(-\omega_{\ell}) \\
\zeta_{2a}(-\omega_{\ell}) \\
\end{array}
\right),\nonumber\\
\end{eqnarray}
where
\begin{eqnarray}
&&M_{\zeta} \nonumber\\
&=& \left(
\begin{array}{cc}
\omega_{\ell}^2N_1(0)^2+\frac{4\pi e^2N(0)^2n_1}{m_{1a}} & \omega_{\ell}^2N_1(0)N_2(0) \\
\omega_{\ell}^2N_1(0)N_2(0) & \omega_{\ell}^2N_2(0)^2+\frac{4\pi e^2N(0)^2n_2}{m_{2a}}\\
\end{array}
\right),\nonumber\\
\end{eqnarray}
where $N(0)=N_1(0)+N_2(0)$.
Then, the plasma frequency is given by the solution of 
${\rm det}M_{\zeta}(i\omega_{\ell}\rightarrow \omega+i\delta)=0$:
\begin{equation}
\omega_{pl,a}^2= 4\pi e^2\frac{n_1n_2}{m_{1a}m_{2a}}
\frac{(N_1(0)+N_2(0))^2}{ \frac{n_1}{m_{1a}}N_2(0)^2+\frac{n_2}{m_{2a}}N_1(0)^2}.
\end{equation}
In an $N$-gap superconductor, the plasma frequency is generalized to be
\begin{eqnarray}
\omega_{pl,a}^2 &=& 4\pi e^2\frac{n_1\cdots n_N}{m_{1a}\cdots m_{Na}}
N(0)^2 \Big[ N_1(0)^2\frac{n_2\cdots n_N}{m_{2a}\cdots m_{Na}}
\nonumber\\
&& +\cdots
+N_N(0)^2\frac{n_1\cdots n_{N-1}}{m_{1a}\cdots m_{N-1,a}} \Big]^{-1},
\nonumber\\
\end{eqnarray}
for $a=x$, $y$ and $z$ where $N(0)=\sum_jN_j(0)$.
When $N$ gaps are equivalent, this formula reduces to
\begin{equation}
\omega_{pl,a}^2= 4\pi e^2\frac{n}{m_a}N.
\end{equation}

\section{Higgs Green's function in superconductors}

\subsection{Higgs Green's function}

The Green functions for the Higgs boson $h_n=\psi_n^{\dag}\sigma_1\psi_n$
are defined by
\begin{equation}
H_{nm}(x-y)= -i\langle Th_n(x)h_m(y)\rangle.
\end{equation}
The effective action for the Higgs fields, up to the one-loop order,
is written as
\begin{eqnarray}
S[h] &=& -\sum_{nm}\int dtd^dx h_n(g^{-1})_{nm}h_m\nonumber\\
&&+\frac{i}{2}\sum_n{\rm Tr}h_nG_n^{(0)}(p)\sigma_1G_n^{(0)}(p+q)
\sigma_1h_n.
\label{haction}
\end{eqnarray}
When the temperature $T$ is near $T_c$, this gives the well-known
time-dependent Ginzburg-Landau (TDGL) action.
In the TDGL action, due to the dissipation effect, the Higgs mode
may not be defined clearly.
The Higgs mode is well defined at low temperatures.

The second term in the effective action eq.(\ref{haction}) for the Higgs field
is the one-loop contribution given by
\begin{eqnarray}
\Pi_n(i\epsilon,{\bf q}) &=& \frac{1}{2\beta}\sum_n\frac{1}{V}
\sum_{{\bf p}}{\rm tr}\Big[ G_n^{(0)}(i\omega_n+i\epsilon,{\bf p}+{\bf q})
\sigma_1 \nonumber\\
&& \times G_n^{(0)}(i\omega_n,{\bf p})\sigma_1\Big],
\end{eqnarray}
where we use the Matsubara formalism.
At absolute zero, $\Pi_n(q_0,{\bf q})$ (where $q_0=i\epsilon$) is
expanded in $|{\bf q}|$ as
\begin{eqnarray}
\Pi_n(q_0,{\bf q}) &=& -N_n(0)\int d\xi\frac{1}{2E_n(\xi)} 
\nonumber\\
&& +N_n(0)\left( 4\bar{\Delta}_n^2-q_0^2\right)\frac{1}{4\bar{\Delta}_n^2}
F\left(\frac{q_0}{2\bar{\Delta}_n}\right)\nonumber\\
&& + N_n(0)\frac{1}{3}c_{ns}^2\left( \frac{{\bf q}}{2\bar{\Delta}_n}\right)^2
+\cdots ,
\end{eqnarray}
where $c_{ns}^2=v_{Fn}^2/3$ and $E_n(\xi)=\sqrt{\xi^2+\bar{\Delta}_n^2}$.
$F(x)$ is given by\cite{koy16}
\begin{eqnarray}
F(x)&=& \frac{1}{x\sqrt{1-x^2}}\tan^{-1}\left( \frac{x}{\sqrt{1-x^2}}\right)
(0\le x<1)\\
F(x)&=& -\frac{i}{x\sqrt{x^2-1}}\frac{\pi}{2}
+\frac{1}{2x\sqrt{x^2-1}}\ln\Big|\frac{x-\sqrt{x^2-1}}{x+\sqrt{x^2-1}}\Big|
\nonumber\\
&& ~~~~~(1<x).
\end{eqnarray} 
We used an approximation that the density of states is constant.

When $q_0$ and ${\bf q}$ are small, we have for a single-band
superconductor
\begin{eqnarray}
\frac{1}{g}+\Pi(q_0,{\bf q})&=&  N(0)\Big[ 1-\frac{1}{3}
\left(\frac{q_0}{2\bar{\Delta}}\right)^2\Big]
+N(0)\frac{1}{3}c_s^2\left(\frac{{\bf q}}{2\bar{\Delta}}\right)^2.
\nonumber\\
\end{eqnarray} 
This is proportional to the inverse of the Fourier transform of
the Higgs Green function.
This agrees with the effective action for the Higgs mode 
obtained by means of the functional integral method given as\cite{yan17,ait99}
\begin{eqnarray}
S^{(2)}[h]&=& \int d\tau d^dx N(0)\Big[ \frac{1}{12\bar{\Delta}^2}
\left(\frac{\partial h}{\partial\tau}\right)^2
+\frac{v_F^2}{36\bar{\Delta}^2}\left(\nabla h\right)^2
\nonumber\\
&& +h^2 \Big],
\label{higgsact2}
\end{eqnarray}
in the imaginary-time formulation.
Please note that $1/g+\Pi(q)$ does not have a zero when $q_0$ is
small as far as $0\le q_0<2\bar{\Delta}$.
We show the behavior of $F(x)$ in Fig.8 and $g^{-1}+\Pi(\omega,{\bf q}=0)$ 
as a function of $\omega$ in Fig.9.

When $v_{Fn}|{\bf q}|/q_0\ll 1$, the $g^{-1}+\Pi$ is given by
\begin{eqnarray}
\left( g^{-1}+\Pi\right)_{mn} &=& (g^{-1})_{mn}-\delta_{mn}f_n
\nonumber\\
&+& \delta_{mn}N_n(0) \Big[ 1-\left(\frac{\tilde{q}_{n}}{2\bar{\Delta}_n}\right)^2
\Big]
F\left(\frac{\tilde{q}_{n}}{2\bar{\Delta}_n}\right),
\nonumber\\
\end{eqnarray}
where we put
\begin{equation}
\tilde{q}_{n}=\sqrt{ q_0^2-\frac{1}{3}v_{Fn}^2{\bf q}^2},
\end{equation}
for the Fermi velocity $v_{Fn}$ in the $n$-th band.

We indicate the Fourier transform of the Higgs Green function $H_{nm}(x-y)$
as $H_{nm}(q)$:  
\begin{equation}
H_{nm}(x)= \int \frac{dp_0}{2\pi}\frac{d^dp}{(2\pi)^d}
e^{ip_0x_0+i{\bf p}\cdot{\bf x}}H_{nm}(p),
\end{equation}
for $p=(p_0,{\bf p})$.
In a similar way as to the NG boson, the Higgs Green
function satisfies\cite{koy16}
\begin{eqnarray}
H_{nm}(q)&=& -i\int \frac{dk_0}{2\pi}\frac{d^dk}{(2\pi)^d}{\rm tr}
\Big[ \delta_{nm}\sigma_1 G_n(k)\sigma_1 G_n(k+q) \nonumber\\
&-& \sum_{\ell}\sigma_1 G_n(k)\Lambda_{H,n\ell}(k.k+q) G_n(k+q)
H_{\ell m}(q)\Big],\nonumber\\
\end{eqnarray}
where we introduced the vertex function $\Lambda_{H,n\ell}$.
We can put this in the form
\begin{equation}
\Lambda_{H,n\ell}= \frac{1}{2}g_{n\ell}\sigma_1.
\end{equation}
Then the matrix of Higgs Green's function $H=(H_{nm})$ is written
as follows.
\begin{equation}
H(q)= 2g^{-1}(g^{-1}+\Pi(q))^{-1}\Pi(q),
\end{equation}
where $\Pi(q)$ is the diagonal matrix with elements $\Pi_n(q)$ ($n=1,\cdots$):
$\Pi(q)={\rm diag}(\Pi_1(q),\Pi_2(q),\cdots)$.

In the single-band case, the Higgs Green function for small $q_0$ and
${\bf q}$ is given by
\begin{equation}
H(q)\simeq 2N(0)\frac{1}{gN(0)}\frac{(2\bar{\Delta})^2}
{(2\bar{\Delta})^2-\frac{1}{3}q_0^2+\frac{1}{3}c_s^2{\bf q}^2}.
\end{equation}
When $q_0$ is as large as $2\bar{\Delta}$, $H(q)$ is approximated as
\begin{equation}
H(q)\approx -\frac{8}{\pi}N(0)\left(\frac{1}{gN(0)}\right)^2
\frac{\bar{\Delta}}{\sqrt{\omega({\bf q})^2-q_0^2}}.
\end{equation}
for $q_0 < \omega({\bf q})$ where
\begin{equation}
\omega({\bf q})= \sqrt{ (2\bar{\Delta})^2+\frac{1}{3}v_F^2{\bf q}^2 }.
\end{equation}
In the latter case, the singularity is given by a square root function.
Thus, $H(q)$ represents a fluctuation mode, not a quasiparticle excitation mode
since the residue at $q_0=\omega({\bf q})$ vanishes.
$H(q)$ is defined on a Riemann surface with a cut from 
$q_0=-\omega({\bf q})$ to $q_0=\omega({\bf q})$ om the real axis.
For $q_0>\omega({\bf q})$, there is the imaginary part representing
the damping effect:
\begin{eqnarray}
H(q)^{-1}&\approx& -\frac{1}{2}g^2N(0)\frac{1}{(2\bar{\Delta})^2}
\left( q_0^2-\omega({\bf q})^2\right) \nonumber\\
&& -i\frac{\pi}{4}g^2N(0)\frac{1}{2\bar{\Delta}}
\sqrt{ q_0^2-\omega({\bf q})^2 },
\end{eqnarray}
when $q_0$ is near $\omega({\bf q})$.
This behavior of the Higgs Green function is similar to that
of the $\sigma$-particle Green function in the Gross-Neveu model\cite{gro74}.
The Higgs mode considered here has strong similarity with the $\sigma$ boson
of the Gross-Neveu model in two dimensions.
In fact, the Green function $G_{\sigma}(p)$, where $p=(p_0,p_1)$, of the
$\sigma$ boson
is given as up to the one-loop order
\begin{equation}
iG_{\sigma}(p)^{-1}= g^2 \frac{N_f}{\pi}\sqrt{\frac{4\bar{\Delta}^2-p^2}{p^2}}
\tan^{-1}\sqrt{ \frac{p^2}{4\bar{\Delta}^2-p^2} },
\end{equation}
for $0<p^2\equiv p_0^2-p_1^2<(2\bar{\Delta})^2$,
and
\begin{eqnarray}
iG_{\sigma}(p)^{-1}&=& -g^2\frac{N_f}{2\pi}\Bigg[ \sqrt{\frac{p^2-4\bar{\Delta}^2}{p^2}}
\ln\left|
\frac{\sqrt{p^2}-\sqrt{p^2-4\bar{\Delta}^2}}{\sqrt{p^2}+\sqrt{p^2-4\bar{\Delta}^2}}
\right| \nonumber\\
&& ~~ +i\pi\sqrt{ \frac{p^2-4\bar{\Delta}^2}{p^2} } \Bigg],
\end{eqnarray}
for $p^2>(2\bar{\Delta})^2$,
where $N_f$ is the number of fermions and $g$ is the coupling constant 
in the Gross-Neveu model.
For $p^2<0$ we have
\begin{equation}
iG_{\sigma}(p)^{-1}= -g^2\frac{N_f}{2\pi}\sqrt{\frac{-p^2+4\bar{\Delta}^2}{-p^2}}
\ln\left|
\frac{\sqrt{-p^2}-\sqrt{-p^2+4\bar{\Delta}^2}}{\sqrt{-p^2}+\sqrt{-p^2+4\bar{\Delta}^2}}
\right|.
\end{equation}

In the two-band case, the Higgs Green function is given as 
\begin{eqnarray}
H(q)&=& 2g^{-1}\left(
\begin{array}{cc}
\gamma_{11}+\Pi_1  &  \gamma_{12}  \\
\gamma_{21}  &  \gamma_{11}+\Pi_2 \\
\end{array}
\right)^{-1}\left(
\begin{array}{cc}
\Pi_1  &  0  \\
0  &  \Pi_2  \\
\end{array}
\right)\nonumber\\
&=& \frac{2}{ (\gamma_{11}-f_1)d_2+(\gamma_{22}-f_2)d_1+d_1d_2}\nonumber\\
&&\times \Big[ \frac{1}{{\rm det}g}\left(
\begin{array}{cc}
\Pi_1  &  0  \\
0  &  \Pi_2  \\
\end{array}
\right)+\Pi_1\Pi_2\left(
\begin{array}{cc}
\gamma_{11}  &  \gamma_{12}  \\
\gamma_{21}  &  \gamma_{22}  \\
\end{array}
\right) \Big],\nonumber\\
\end{eqnarray}
where we set $\Pi_n(q)=-f_n+d_n(q)$ where
\begin{equation}
d_n(q)= N_n(0)\Big[ 1-\left(\frac{\tilde{q}_n}{2\bar{\Delta}_n}\right)^2
\Big]F\left(\frac{\tilde{q}_n}{2\bar{\Delta}_n}\right),
\end{equation}
and $\gamma_{nm}= (g^{-1})_{nm}$.
We assume that $\gamma_{nm}=\gamma_{mn}$.
Here we used the gap equation
\begin{eqnarray}
{\rm det}\left(
\begin{array}{cc}
\gamma_{11}-f_1  &  \gamma_{12}  \\
\gamma_{21}  &  \gamma_{22}-f_2  \\
\end{array}
\right)=0.
\end{eqnarray}
It is seen from this expression that the Green function $H(q)$ shows no
divergence when $\bar{\Delta}_1\neq \bar{\Delta}_2$ in general.
This is shown in Fig. 10.
The Higgs mode is not a quasiparticle mode and  exists as a fluctuation mode 
in a multiband superconductor.

We define the Higgs constant potential $V=(V_{mn})$ as
\begin{equation}
V_{mn}= (g^{-1})_{mn}+\delta_{mn}\Pi_n(q_0=0,{\bf q}=0). 
\end{equation}
This is something like the 'mass term' for Higgs fields $h_n$.
In the subsection 5.4,
we evaluate the eigenvalue $y_H^2$ of this matrix, by solving
the gap equation.
In a usual field theory, $y_H$ indicates the mass of the field $h$.
In a superconductor, however, $y_H$ is different from the excitation 
gap because
$y_H$ is not determined from a singularity of the Higgs Green function.
The eigenvalues $y_{Hn}$, however, characterize the Higgs modes and determine 
the spatial expanse of the Higgs fields. 
In case ${\rm det}(V_{mn})=0$, one eigenvalue vanishes.

\begin{figure}[htbp]
\begin{center}
\includegraphics[width=7.0cm]{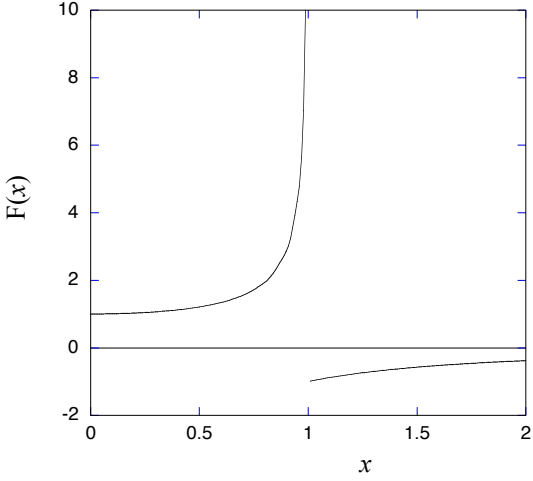}
\caption{$F(x)$ as a function of $x=q_0/2\bar{\Delta}$.
$F(x)$ has a singularity at $x=1$.
The 'mass' term is obtained by expanding $(1-x^2)F(x)$ in
terms of $x$.
}
\label{Fx}
\end{center}
\end{figure}

\begin{figure}[htbp]
\begin{center}
\includegraphics[width=7.0cm]{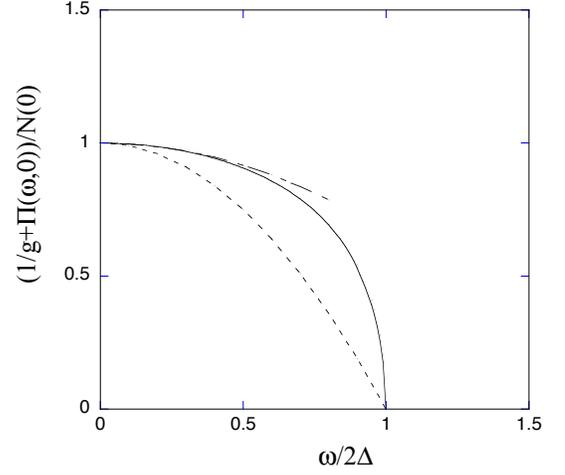}
\caption{$g^{-1}+\Pi(\omega,{\bf q}=0)$ as a function of the
$x\equiv \omega/2\bar{\Delta}$.
The dash-dotted line shows the curve $1-x^2/3$.
The dotted line indicates $1-x^2$ that corresponds to the
Nambu-Jona-Lasinio model where $N(0)$ is replaced by the
divergent integral $I_0\equiv (1/4\pi^2)\ln(\Lambda/\bar{\Delta})$
with the cutoff $\Lambda$.
}
\label{gPi}
\end{center}
\end{figure}

\begin{figure}[htbp]
\begin{center}
\includegraphics[width=7.5cm]{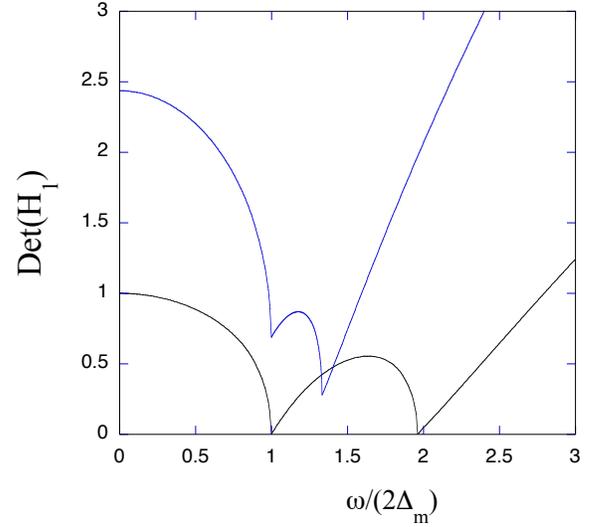}
\caption{${\rm det}H_1$ as a function of $\omega/2\Delta_m$
in a two-band case where $H_1={\rm Re}(g^{-1}+\Pi)$.
The lower curve is for $N_1g_{11}=0.3$, $N_2g_{22}=0.25$ and
$N_1g_{12}=0$ and the upper curve is for $N_1g_{11}=0.3$,
$N_2g_{22}=0.25$ and $N_1g_{12}=0.05$.
We put $\Delta_m=min(\bar{\Delta}_1,\bar{\Delta}_2,\bar{\Delta}_3)$.
When the interband $g_{12}$ vanishes, we have zeros at $\omega=\bar{\Delta}_1$
and $\omega=\bar{\Delta}_2$.
}
\label{detH}
\end{center}
\end{figure}

\begin{figure}[htbp]
\begin{center}
\includegraphics[width=7.0cm]{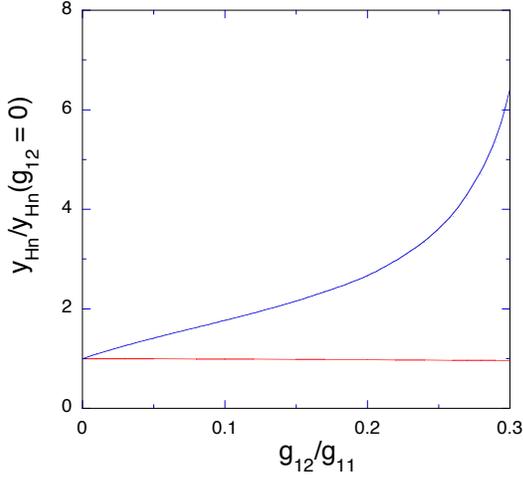}
\caption{Eigenvalue $y_H^2$ in a two-band superconductor as a function of the
interband coupling $g_{12}$.
$y_H$ is measured in unit of that in the case of $g_{12}=0$.
We set $g_{11}\bar{N}(0)=g_{22}\bar{N}(0)=0.35$ for $\bar{N}(0)=N_1(0)=N_2(0)$.
The Higgs 'mass' of one mode remains a constant whereas the other 'mass'
grows very large as $g_{12}$ increases.
}
\label{mH}
\end{center}
\end{figure}

\subsection{Kinetic terms of the Higgs mode}

We discuss the kinetic terms of Higgs boson in this subsection briefly, 
where the kinetic terms mean $(\partial_{\tau}h)^2$ and
$(\nabla h)^2$.  
The time-dependent Ginzburg-Landau model with dissipation is often used
in a study of nonequilibrium properties of superconductors near
the critical temperature $T_c$.
At low temperature $T\ll T_c$, the action for the Higgs mode is
given by the quadratic from as shown in the previous subsection.
Thus there is a temperature dependence.
We consider here the single-band case for simplicity.
The coefficient of $h\partial_{\tau}^2 h$ is given as, up to the
one-loop order,
\begin{eqnarray}
I_{\tau}\equiv \frac{1}{\beta}\sum_{\ell}\frac{1}{V}\sum_{{\bf k}}&\Big[& 
\frac{\bar{\Delta}^2-\omega_{\ell}^2-\xi_k^2}
{(\omega_{\ell}^2+\xi_k^2+\bar{\Delta}^2)^3}
-2\frac{\omega_{\ell}^2}{(\omega_{\ell}^2+\xi_k^2+\bar{\Delta}^2)^3}
\nonumber\\
&& -4\frac{\bar{\Delta}^2-\omega_{\ell}^2-\xi_k^2}
{(\omega_{\ell}^2+\xi_k^2+\bar{\Delta}^2)^4}\omega_{\ell}^2 \Big],
\end{eqnarray}
where $\xi_k$ denotes the electron dispersion relation.
The coefficient of $h({\bf v}\cdot\nabla)^2 h$ is similarly
give by
\begin{eqnarray}
I_x\equiv \frac{1}{\beta}\sum_{\ell}\frac{1}{V}\sum_{{\bf k}}&\Big[& 
\frac{\bar{\Delta}^2-\omega_{\ell}^2-\xi_k^2}
{(\omega_{\ell}^2+\xi_k^2+\bar{\Delta}^2)^3}
-2\frac{\xi_k^2}{(\omega_{\ell}^2+\xi_k^2+\bar{\Delta}^2)^3}
\nonumber\\
&& -4\frac{\bar{\Delta}^2-\omega_{\ell}^2-\xi_k^2}
{(\omega_{\ell}^2+\xi_k^2+\bar{\Delta}^2)^4}\xi_k^2 \Big].
\end{eqnarray}
At absolute zero, $I_{\tau}$ and $I_x$ are the same since we can exchange
$\omega_{\ell}$ and $\xi_k$.
When $T$ approaches $T_c$, $I_{\tau}$ vanishes at some $T$.
Then, when $T\sim T_c$, the time-dependence of $h$ field should be
given by the time-dependent Ginzburg-Landau functional.
We show the kinetic terms at $T=0$ and $T\sim T_c$ in Table II.

\begin{table}
\caption{Kinetic terms of the Higgs field $h$ in the action at absolute zero
$T\sim 0$ and near $T\sim T_c$ (for the single-band case).
}
\begin{center}
\begin{tabular}{|c|c|c|}
\hline
   & $T\sim 0$ & $T\sim T_c$ \rule[0mm]{0mm}{3mm} \\
\hline
$\tau$-part & $\frac{1}{12\bar{\Delta}^2}N(0)
\left(\frac{\partial h}{\partial\tau}\right)^2$ &
$\frac{\pi\hbar}{8k_BT_c}
hi\frac{\partial h}{\partial\tau}$ \rule[0mm]{0mm}{5mm} \\
$x$-part &
$\frac{1}{36\bar{\Delta}^2}N(0)\hbar^2v_F^2
\left(\nabla h\right)^2$ &
$\frac{7\zeta(3)\hbar^2v_F^2}{48\pi^2k_B^2T_c^2}N(0)
\left(\nabla h\right)^2$ \rule[0mm]{0mm}{5mm} \\
\hline
\end{tabular}
\end{center}
\end{table}

\subsection{Higgs constant potential}

The Higgs constant potential $V$ is defined by $V=g^{-1}+\Pi$ in the limit
of $q_0\rightarrow 0$ and ${\bf q}\rightarrow 0$.
The potential $V$ in a multi-band superconductor is crucially dependent
upon the coupling-constant matrix $g$.
The quadratic from of $h_n$ in this limit is given as
\begin{eqnarray}
M[h]&\equiv& \sum_{mn}h_mM_{mn}h_n \nonumber\\
&=& \sum_{mn}h_m\left( \gamma_{mn}-\delta_{mn}f_m+\delta_{mn}\rho_m\right)
h_n,
\end{eqnarray}
where we set
\begin{eqnarray}
\gamma_{mn} &=& (g^{-1})_{mn}, \\
\rho_m &=& \int \frac{d^dk}{(2\pi)^d}\frac{1}{\beta}\sum_{\ell}
\frac{2\bar{\Delta}_m^2}{(\omega_{\ell}^2+\xi_k^2+\bar{\Delta}_m^2)^2}.
\end{eqnarray}
$\gamma_{mn}$ indicates the strength of the Josephson coupling between
$m$ and $n$ bands.
$\rho_m=\rho_m(T)$ equals the density of states $N_m(0)$ at absolute zero and
is proportional to $\bar{\Delta}_m^2$ near $T_c$:
\begin{eqnarray}
\rho_m= \left\{
\begin{array}{cc}
N_m(0) & {\rm at}~ T=0, \\
N_m(0)\frac{7\zeta(3)}{4\pi^2k_B^2T_c^2}\bar{\Delta}_m(T)^2 & {\rm for}~T\sim T_c. \\
\end{array}
\right.
\end{eqnarray}
The gap function $\bar{\Delta}_m$ are determined by the gap equation,
\begin{equation}
\sum_m\gamma_{nm}\bar{\Delta}_m = f_n\bar{\Delta}_n.
\end{equation}

In the single-band case, we have $M_{11}=\rho_1$ because of the gap
equation.
In the two-band case, $M= (M_{mn})$ is given by the $2\times 2$ matrix:
\begin{eqnarray}
M= \left(
\begin{array}{cc}
\gamma_{11}-f_1+\rho_1  &  \gamma_{12} \\
\gamma_{21}  &  \gamma_{22}-f_2+\rho_2  \\
\end{array}
\right).
\end{eqnarray}
The critical temperature $T_c$ is given as
\begin{equation}
k_BT_c = \frac{2e^{\gamma}\omega_c}{\pi} e^{-\lambda},
\end{equation}
where
\begin{eqnarray}
\lambda &=& \frac{g_{11}/N_2+g_{22}/N_1}{2{\rm det}g}
-\frac{1}{{\rm det}g}\sqrt{ \frac{1}{4}\left(\frac{g_{11}}{N_2}
-\frac{g_{22}}{N_1}\right)^2+\frac{g_{12}g_{21}}{N_1N_2} },
\nonumber\\
\end{eqnarray}
and $\omega_c$ is the cutoff.
In the simple case where two bands are equivalent with $g_{11}=g_{22}$,
$g_{12}=g_{21}$
and $N_1=N_2$, the superconducting gap at $T=0$ is
\begin{equation}
\bar{\Delta}_1 = \bar{\Delta}_2 = 2\omega_c e^{-\lambda},
\end{equation}
where
\begin{equation}
\lambda = \frac{1}{{\rm det}g}\left( \frac{g_{11}}{N_1}-\frac{|g_{12}|}{N_1}
\right) = \frac{1}{(g_{11}+|g_{12}|)N_1}.
\end{equation}
In this case, we must notice that $\Delta_n$ and $T_c$ are finite
even when ${\det}g=0$.

When ${\rm det}g=0$, the gap functions are obtained as
\begin{eqnarray}
\bar{\Delta}_1 &=& 2\omega_c\exp\left( -\frac{1}{g_{11}N_1+g_{22}N_2}
\left(1-\frac{1}{2}g_{22}N_2\ln\left(\frac{g_{22}}{g_{11}}\right)\right)\right),
\nonumber\\
\end{eqnarray}
for $g_{12}=g_{21}$.  $\bar{\Delta}_2$ is obtained by exchanging
the indices 1 and 2.
This is in contrast to the single-band case where the vanishing
of $g$ means the disappearance of superconductivity.
A singular behavior of the 'mass' spectra occurs when the determinant
${\rm det}g$ is small.
We call the region with small ${\rm det}g$ the critical region in
the following.
We find that one of the eigenvalues of the matrix $M$ can be
very large in this region.

\subsection{Spectra of the Higgs potential in the two-band model}

Let us consider the two-band case.
There are two cases that should be examined; they are (a) ${\rm det}g>0$
and (b) ${\rm det}g<0$.
Since the fields $h_i$ have the same dimension as $\Delta_i$, 
we define the fields $\eta_n$ by
\begin{equation}
h_n = \bar{\Delta}_n\eta_n.
\end{equation}
Correspondingly, we define the Higgs mass matrix $\tilde{M}$ as
\begin{eqnarray}
\tilde{M} = \left(
\begin{array}{cc} 
(\gamma_{11}-f_1+\rho_1)\bar{\Delta}_1^2 & \gamma_{12}\bar{\Delta}_1\bar{\Delta}_2 \\
\gamma_{12}\bar{\Delta}_1\bar{\Delta}_2 & (\gamma_{22}-f_2+\rho_2)\bar{\Delta}_2^2 \\
\end{array}
\right).\nonumber\\
\end{eqnarray}
\\
(a) Higgs modes for ${\rm det}g>0$.
\\
\\
In this case, ${\rm det}g= g_{11}g_{22}-g_{12}g_{21}>0$.
The Josephson couplings $\gamma_{ij}$ are
\begin{eqnarray}
\gamma_{11}&=&\frac{g_{22}}{{\rm det}g},~~ \gamma_{22}=\frac{g_{11}}{{\rm det}g},
~~ \gamma_{12}=-\frac{g_{12}}{{\rm det}g}.
\nonumber\\
\end{eqnarray}
We assume that $g_{11}>0$ and $g_{22}>0$, namely, the interaction
in each band is attractive.
Then, w have $\gamma_{11}>0$ and $\gamma_{22}>0$.
We also set $g_{12}=g_{21}$.
When $\gamma_{12}=0$, the gap equation reduces to
$\gamma_{nn}-f_n=0$ ($n=1,2$).  For $\gamma_{12}\neq 0$, $f_n$ decreases,
that is, $\bar{\Delta}_n$ increases, and thus $\gamma_{nn}-f_n>0$.  
The eigenvalues $x$ of $\tilde{M}$ are given as
\begin{eqnarray}
x &=& \frac{1}{2}\{(\gamma_{11}-f_1+\rho_1)\bar{\Delta}_1^2
+(\gamma_{22}-f_2+\rho_2)\bar{\Delta}_2^2\}\nonumber\\
&& \pm\frac{1}{2}\Big[ (\gamma_{11}-f_1+\rho_1)\bar{\Delta}_1^2
-(\gamma_{22}-f_2+\rho_2)\bar{\Delta}_2^2 \nonumber\\
&&+4\gamma_{12}^2\bar{\Delta}_1^2\bar{\Delta}_2^2 \Big]^{1/2}.
\end{eqnarray}

Let us consider, for simplicity, a simple model with equivalent two bands
satisfying $g_{11}=g_{22}$, $N_1(0)=N_2(0)\equiv \bar{N}(0)$ and
$\bar{\Delta}_1=\bar{\Delta}_2\equiv \bar{\Delta}$.
This leads to $f_1=f_2$, $\rho_1=\rho_2$ and $\gamma_{11}-f_1=|\gamma_{12}|$.
The eigenvalues of $\bar{M}$ are
\begin{equation}
x_1= \rho_1\bar{\Delta}^2,~~x_2= (\rho_1+2|\gamma_{12}|)\bar{\Delta}^2.
\end{equation}
The corresponding $y_{H1}$ and $y_{H2}$ are given by
\begin{equation}
y_{H1}= \sqrt{\frac{\rho_1}{\bar{N}(0)}}\bar{\Delta},~~
y_{H2}= \sqrt{ \frac{\rho_1+2|\gamma_{12}|}{\bar{N}(0)} }\bar{\Delta}.
\end{equation}
The one mode $y_{H2}$ shows a dependence on $g_{12}$, while the
other value remains a constant.
$y_{H2}$ increases as $|g_{12}|$ increases as shown in Fig.11 for the
two equivalent bands, with a
very large value when ${\rm det}g$ is small.
The Fig.12 indicates the superconducting gaps and $y_H$'s for
$g_{11}N_1(0)= 0.35$ and $g_{22}N_2(0)= 0.30$ at absolute zero.
The coherence length $\xi$ is proportional to the inverse of $y_H$,
exhibiting the dependence on the coupling constant matrix $g$.
Thus the upper critical field $H_{c2}$, being proportional to
$1/\xi^2$, also shows the dependence on $g$.
$H_{c2}$ can be very large as ${\rm det}g$ approaches zero.
\\
\\
(b) Higgs potential for ${\rm det}g<0$.
\\
\\
Let us turn to the case with ${\rm det}g=g_{11}g_{22}-g_{12}g_{21}<0$.
We assume that $g_{11}>0$ and $g_{22}>0$.
This means that $\gamma_{11}<0$ and $\gamma_{22}<0$.
We examine the case with two equivalent bands.
Since $\gamma_{11}<0$, we have $\gamma_{11}-f_1=-|\gamma_{12}|$.
Then, the eigenvalues of the matrix $\bar{M}$ are
\begin{equation}
x_1= \rho_1\bar{\Delta}^2,~~x_2= (\rho_1-2|\gamma_{12}|)\bar{\Delta}^2.
\end{equation}
Correspondingly, we have
\begin{equation}
y_{H1}= \sqrt{\frac{\rho_1}{\bar{N}(0)}}\bar{\Delta},~~
y_{H2}= \sqrt{ \frac{\rho_1-2|\gamma_{12}|}{\bar{N}(0)} }\bar{\Delta}.
\end{equation}
In constrast to the previous case, $y_{H2}$ decreases as $|\gamma_{12}|$
increases.  Thus, the coherence length corresponding to $y_{H2}$ increases
and can be very large as a function of $g_{12}|$. 
$y_{H2}$ decreases and vanishes when $|\gamma_{12}|$ increases as
$|g_{12}|$ approaches $g_{11}$.
The appearance of massless state indicates an instability of the superconducting
state.  When $y_{H2}^2<0$, the state with 
$(\bar{\Delta}_1,\bar{\Delta}_2)=(\bar{\Delta},\bar{\Delta})$ is at the saddle
point and thus is instable to be away from this point.  
Let us examine this phenomenon in the following.

We include $\eta^4$ term in the action to investigate a stability of 
superconducting state.
The mass functional is written as
\begin{equation}
M[\eta]= a_1\eta_1+a_2\eta_2+\eta^t\tilde{M}\eta+\frac{b}{4}(\eta_1^4+\eta_2^4),
\end{equation}
with a constant $b>0$, where $\eta= ^{t}(\eta_1,\eta_2)$.
$\bar{\Delta}_n$ ($n=1,2$) are determined from the condition
$a_1=a_2=0$.
We write $\eta_n= \bar{\eta}_n+\delta\eta_n$ (for $n=1,2$) where 
$\bar{\eta}_n$ is a new stationary value of $\eta_n$ and $\delta\eta_n$
stands for fluctuation of  the mode $\eta_n$.
The condition that the linear terms in $\delta\eta_n$ should vanish
results in the equations
\begin{eqnarray}
2\tilde{M}_{11}\bar{\eta}_1+2\tilde{M}_{12}\bar{\eta}_2+b\bar{\eta}_1^3 &=& 0,\\
2\tilde{M}_{12}\bar{\eta}_1+2\tilde{M}_{22}\bar{\eta}_2+b\bar{\eta}_2^3 &=& 0.
\end{eqnarray}
\\
Let us consider the model with two equivalent bands where we set
$\bar{\Delta}\equiv \bar{\Delta}_1=\bar{\Delta}_2$ and $g_{11}=g_{22}$.
The eigenvectors of the matrix $\bar{M}$ are
$^t(1,1)/\sqrt{2}$ and $^t(1,-1)/\sqrt{2}$ with the eigenvalue 
$\rho_1\bar{\Delta}^2$ and $(\rho_1-2|\gamma_{12}|)\bar{\Delta}^2$,
respectively.
The eigenvalue corresponding to the mode $\eta_1-\eta_2$ can be negative
and the state will show an instability to this direction.
Then, we set $\bar{\eta}_2=-\bar{\eta}_1$ to obtain the equation
\begin{equation}
2(\rho_1-2|\gamma_{12}|)\bar{\Delta}^2+b\bar{\eta}_1^2=0,
\end{equation}
for non-trivial solution $\bar{\eta}_1\neq 0$.  Then we have
\begin{equation}
\bar{\eta}_1= \pm \sqrt{ \frac{2(2|\gamma_{12}|-\rho_1)}{b} }
\bar{\Delta}.
\end{equation}
Due to fluctuation of the $\eta_1-\eta_2$ mode,
the stationary values of the gap functions $(\bar{\Delta}_1,\bar{\Delta}_2)$
change from $(\bar{\Delta},\bar{\Delta})$ to
\begin{equation}
(\bar{\Delta}+\bar{\eta}_1\bar{\Delta},\bar{\Delta}-\bar{\eta}_1\bar{\Delta}).
\end{equation}
This is shown schematically in Fig. 13.
Thus, the stationary values of $\Delta_n$ break the symmetry
$\Delta_1=\Delta_2$ which should hole for the case of two equivalent bands,
when ${\rm det}g<0$ and $\rho_1-2|\gamma_{12}|<0$.
This may be called the spontaneous symmetry breaking of ${\bf Z}_2$
symmetry.
 
Near $( (1+\bar{\eta}_1)\bar{\Delta},(1-\bar{\eta}_1)\bar{\Delta} )$,
the potential $M[\eta]$ is expanded as
\begin{eqnarray}
M[\eta]= {\rm const.}+^{t}\delta\eta \left(
\begin{array}{cc}
\tilde{M}_{11}+\frac{3}{2}b\bar{\eta}_1^2 & \tilde{M}_{12} \\
\tilde{M}_{21} & \tilde{M}_{22}+\frac{3}{2}b\bar{\eta}_2^2 \\
\end{array}
\right) \delta\eta,
\nonumber\\
\end{eqnarray}
where $\delta\eta= ^{t}(\delta\eta_1,\delta\eta_2)$.
The eigenvalues of this $2\times 2$ matrix give $y_H$'s for
Higgs modes.  For the case with two equivalent bands, we have
\begin{equation}
y_{H1}^2=\frac{2(3|\gamma_{12}|-\rho_1)}{\bar{N}(0)}\bar{\Delta}^2,~~
y_{H2}^2= \frac{2(2|\gamma_{12}|-\rho_1)}{\bar{N}(0)}\bar{\Delta}^2.
\end{equation}
This is shown in Fig.14 where $y_{Hn}^2$ for $n=1,2$ are shown
as a function of $|\gamma_{12}|/\rho_1$.
One mode becomes massless when $2|\gamma_{12}|=\rho_1$.
When $2|\gamma_{12}|>\rho_1$, namely, $g_{12}^2$ is near $g_{11}g_{22}$,
the Higgs values $y_H$ can be very large.
 
As we have shown above, the eigenvalues of Higgs matrix $M$ exhibit a
singular behavior when ${\rm det}g$ is small.
When ${\rm det}g>0$, the eigenvalue and thus $H_{c2}$ can be large.
When ${\rm det}g<0$, there is a possibility of softening of the
eigenvalue of one Higgs mode.
We show $y_{Hn}^2$ as a function of $g_{12}$ in Fig.15.
There is a singularity in the critical region near ${\rm det}g= 0$,
which shows a possibility of large upper critical field $H_{c2}$. 

\begin{figure}[htbp]
\begin{center}
\includegraphics[width=8.0cm]{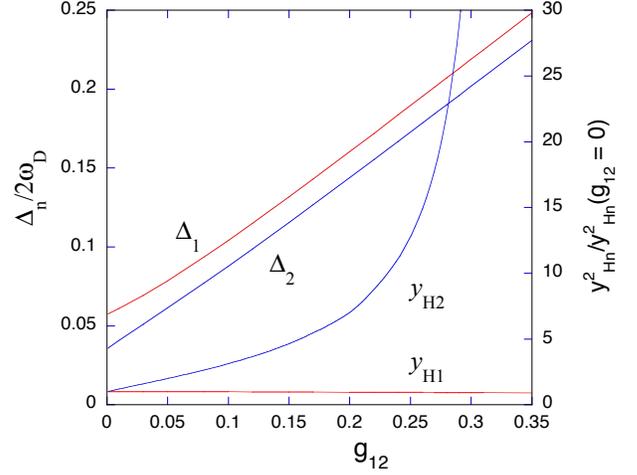}
\caption{Gap functions  $\bar{\Delta}_n$ and Higgs eigenvalue $y_{Hn}$ in a 
two-band superconductor ($n=1,2$)
as a function of the
interband coupling $g_{12}$.
$y_{Hn}$ is measured in unit of that in the case of $g_{12}=0$.
We set $g_{11}\bar{N}(0)=0.35$ and $g_{22}\bar{N}(0)=0.30$ for 
$\bar{N}(0)=N_1(0)=N_2(0)$.
The Higgs eigenvalue of one mode remains a constant whereas the other 
grows very large as $g_{12}$ increases, namely ${\rm det}g$
approaches zero.  $\Delta_n$ in the figure indicates $\bar{\Delta}_n$.
}
\label{mH2}
\end{center}
\end{figure}

\vspace{0.5cm}

\begin{figure}[htbp]
\begin{center}
\includegraphics[width=6.3cm,angle=90]{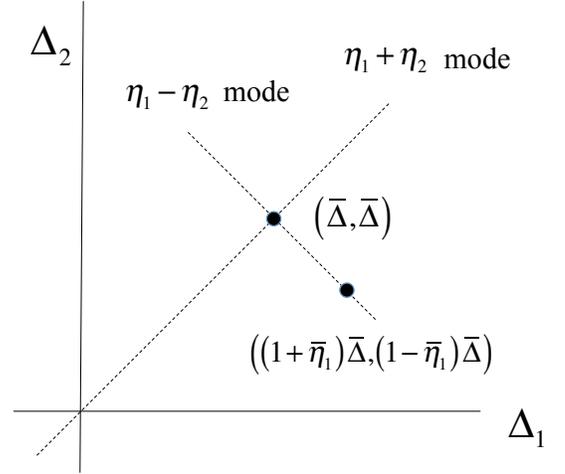}
\caption{Gap functions and fluctuation modes when two bands
are equivalent for ${\rm det}g<0$ and $\rho_1-2|\gamma_{12}|<0$.
The order parameters $(\bar{\Delta},\bar{\Delta})$ is shifted to
$((1+\bar{\eta}_1)\bar{\Delta},(1-\bar{\eta}_1)\bar{\Delta})$ or
$((1-\bar{\eta}_1)\bar{\Delta},(1+\bar{\eta}_1)\bar{\Delta})$
due to the fluctuation of $\eta_1-\eta_2$ mode.
}
\label{eta}
\end{center}
\end{figure}

\begin{figure}[htbp]
\begin{center}
\includegraphics[width=7.0cm]{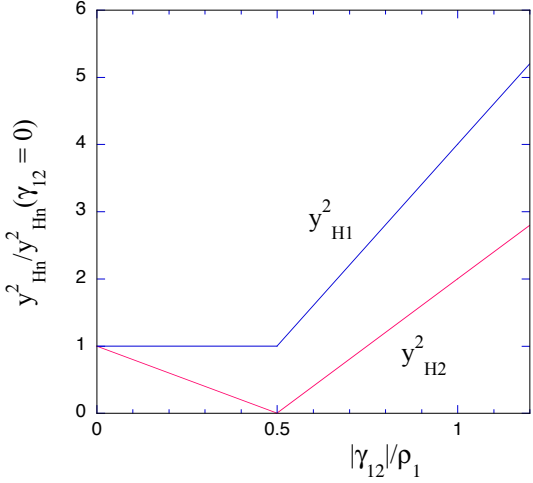}
\caption{$y_{Hn}^2$ ($n=1,2$) as a function of $|\gamma_{12}|$
when ${\det}g<0$.
In this case, the region with small $\gamma_{12}$ is unrealistic
because $g_{12}$ should be very small. 
We used $g_{11}N_1(0)=0.30$.
}
\label{mHgm}
\end{center}
\end{figure}

\begin{figure}[htbp]
\begin{center}
\includegraphics[width=7.5cm]{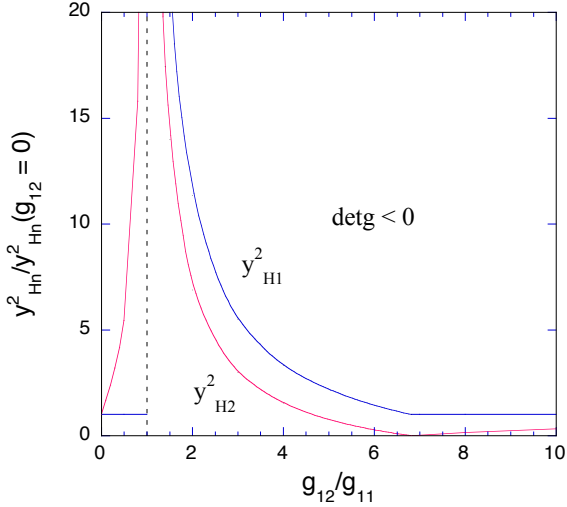}
\caption{$y_{Hn}^2$ ($n=1,2$) as a function of $g_{12}/g_{11}$
where $g_{12}>0$.
We have ${\det}g<0$ for $g_{12}/g_{11}>1$.
$y_{Hn}$ shows a singularity near ${\det}g=0$.
}
\label{mHg12}
\end{center}
\end{figure}

\subsection{Higgs potential in a three-band superconductor}

We turn to a three-band superconductor in this subsection.
The Higgs matrix $M$ is given by
\begin{eqnarray}
M=\left(
\begin{array}{ccc}
\gamma_{11}-f_1+\rho_1 & \gamma_{12} & \gamma_{13} \\
\gamma_{21} & \gamma_{22}-f_2+\rho_2 & \gamma_{33} \\
\gamma_{31} & \gamma_{32} & \gamma_{33}-f_3+\rho_3 \\
\end{array}
\right).\nonumber\\
\end{eqnarray}
Let us first consider the case where three bands are equivalent,
that is, we have $g_{11}=g_{22}=g_{33}\equiv t>0$ and
$g_{12}=g_{21}=g_{23}=g_{32}=g_{31}=g_{13}\equiv v$.
We have also the same density of state in every band as
$N_1(0)=N_2(0)=N_3(0)\equiv \bar{N}(0)$ and $\rho_1=\rho_2=\rho_3\equiv \rho$,
and thus $f_1=f_2=f_3\equiv f$.
In this case the matrix $g$ of coupling constants is
\begin{eqnarray}
g= \left(
\begin{array}{ccc}
t & v & v \\
v & t & v \\
v & v & t \\
\end{array}
\right).
\end{eqnarray}
Then, ${\rm det}g= (t-v)^2(t+2v)$ and the Josephson couplings
$\gamma_{ij}$ are
\begin{equation}
\gamma_{ii}=\frac{t+v}{(t-v)(t+2v)},~~
\gamma_{ij}=-\frac{v}{(t-v)(t+2v)}~(i\neq j).
\end{equation}
The gap equation is written as
\begin{eqnarray}
\left|
\begin{array}{ccc}
\gamma_{11}-f_1 & \gamma_{12} & \gamma_{13} \\
\gamma_{21} & \gamma_{22}-f_1 & \gamma_{23} \\
\gamma_{31} & \gamma_{32} & \gamma_{33}-f_3 \\
\end{array}
\right| = 0.
\end{eqnarray}
In our simple case, the critical temperature is given by
\begin{equation}
k_B T_c = \frac{2e^{\gamma_E}}{\pi}\omega_D\exp\left( 
-\frac{1}{(g_{11}+2g_{12})\bar{N}(0)}\right),
\end{equation}
when $v=g_{12}>0$, and
\begin{equation}
k_B T_c = \frac{2e^{\gamma_E}}{\pi}\omega_D\exp\left( 
-\frac{1}{(g_{11}+|g_{12}|)\bar{N}(0)}\right),
\end{equation}
when $v=g_{12}<0$.
The eigenvalues of the matrix $M$ are 
\begin{eqnarray}
{\rm (A)}~~~x_1&=& \rho,~~x_2= \rho-2\gamma_{12}~~{\rm for}~v>0,\\
{\rm (B)}~~~x_1&=& \rho,~~x_2= \rho+3\gamma_{12}~~{\rm for}~v<0.
\end{eqnarray}
For $v>0$, the eigenstates with $\rho-2\gamma_{12}$ are doubly
degenerate.  On the other hand, for $v<0$, the states with the
eigenvalue $\rho$ are doubly degenerate.
We call these two cases the case A and B, respectively.
In the case A, since $\gamma_{12}$ is negative for $0<v<g_{11}$,
two eigenvalues of the Higgs matrix $M$ increase as $|\gamma_{12}|$ increases.
$\rho-3\gamma_{12}$ diverges at the critical point ${\rm det}g=0$.
In contrast, in the case B, one Higgs mode has a large $y_H$ in the
critical region where ${\rm det}g$ is small.

We present a typical behavior of $y_{Hn}$ as a function of
$g_{12}$ for the case with three equivalent bands in Fig.16.
The 'mass' of one mode remains constant as in the case of two-band
superconductivity.
We show $y_{Hn}$ for two cases as a function of $g_{12}$ in
Figs.17 and 18 where three bands are not necessarily equivalent.
In Fig.17 $y_{Hn}$ of two modes show a large dependence of $g_{12}$
indicating that this is in the case A.

\begin{figure}[htbp]
\begin{center}
\includegraphics[width=7.3cm]{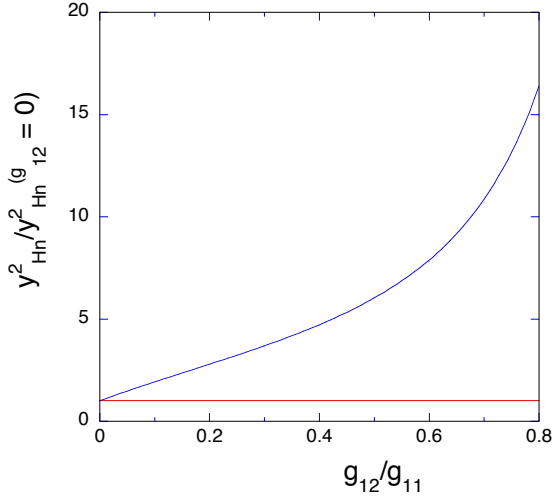}
\caption{$y_{Hn}^2$ ($n=1,2,3$) as a function of $g_{12}/g_{11}$
for the model with three equivalent bands
where we assume $g_{11}=g_{22}=g_{33}$, $g_{12}=g_{23}=g_{31}$ and 
$N_1(0)=N_2(0)=N_3(0)$.  We set $g_{11}N_1(0)=0.3$.
Two Higgs values grows large as $g_{12}(>0)$ increases in the case A.
In the case B, the label of the horizontal axis should read $2|g_{12}|/g_{11}$
for $g_{12}<0$ and we have a similar behavior.
One of Higgs modes has a heavy $y_H$ for large $|g_{12}|/g_{11}$ in 
the case A.
The heavy eigenvalue becomes very huge in the critical region where 
${\rm det}g\sim 0$.
}
\label{mH3gap0}
\end{center}
\end{figure}

\begin{figure}[htbp]
\begin{center}
\includegraphics[width=7.5cm]{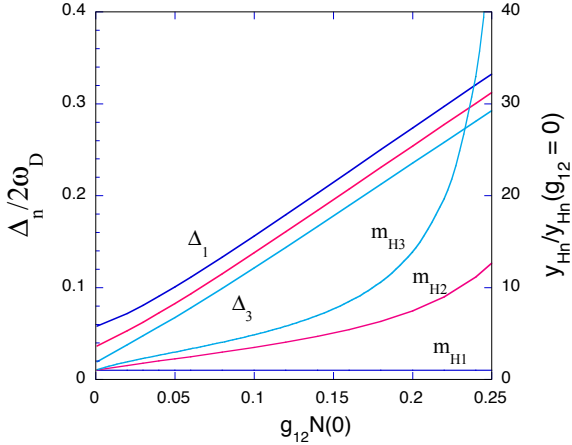}
\caption{$\bar{\Delta}_n$ and $y_{Hn}^2$ ($n=1,2,3$) as a function of $g_{12}/g_{11}$
where we assume $g_{12}=g_{23}=g_{31}$.
We set $g_{11}N_1(0)=0.35$, $g_{22}N_2(0)=0.3$ and $g_{33}N_3(0)=0.25$.
We adopt that we have the same density of states in three bands.
Two of Higgs modes have a heavy value.
}
\label{mH3gap1}
\end{center}
\end{figure}

\begin{figure}[htbp]
\begin{center}
\includegraphics[width=6.8cm]{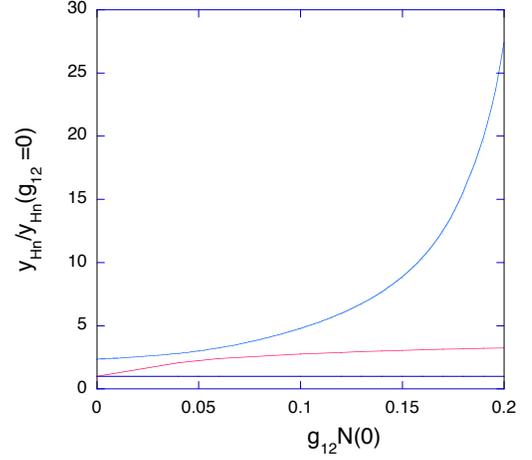}
\caption{$y_{Hn}^2$ ($n=1,2,3$) as a function of $g_{12}\bar{N}(0)$
where we assume $g_{12}=g_{31}$.
We used a set parameters as $g_{11}\bar{N}(0)=0.35$, $g_{22}\bar{N}(0)=0.30$,
$g_{33}\bar{N}(0)=0.25$ and $g_{23}\bar{N}(0)=0.05$.
We also adopt that $N_1(0)=N_2(0)=N_3(0)=\bar{N}(0)$.
In this case, the value of $y_H$ of one Higgs mode becomes large as 
$g_{12}$ increases.
}
\label{mH3gap2}
\end{center}
\end{figure}

Let us investigate the case where $y_{Hn}^2$ is negative 
when $|g_{12}|$ becomes large across the critical point ${\rm det}g=0$.
In the case B with isotropic three bands, the one mode has negative
$y_{H1}^2$ for $v=g_{12}<-t/2=-g_{11}/2$.
The mode $\eta\equiv (\eta_1+\eta_2+\eta_3)/\sqrt{3}$ shows an
instability in this case.  We must include the $(b/4)\sum_i\eta_i^4$ ($b>0$) in
the mass functional $M[\eta_i]$ to examine a stability of
superconducting state.
We express the shift of the stationary point of the gap functions
as $\eta_n= \bar{\eta}+\delta\eta_n$.  We obtain
\begin{equation}
\bar{\eta}= \pm \sqrt{ \frac{2|\rho+3\gamma_{12}|}{b} },
\end{equation}
for $\rho+3\gamma_{12}<0$.  Then, the Higgs matrix $M$ is
\begin{eqnarray}
M= \left(
\begin{array}{ccc}
8|\gamma|-2\rho & \gamma & \gamma  \\
\gamma & 8|\gamma|-2\rho & \gamma \\
\gamma & \gamma & 8|\gamma|-2\rho \\
\end{array}
\right),
\end{eqnarray}
where $\gamma=\gamma_{12}$.  The eigenvalues are
\begin{equation}
x_1= 9|\gamma|-2\rho,~~ x_2= 6|\gamma|-2\rho,
\end{equation}
where the eigenstates with the eigenvalue $x_1=9|\gamma|-2\rho$ are doubly
degenerate.
The Higgs value squared $y_{Hn}^2\propto x_n$ is shown as a function of
$\gamma=\gamma_{12}$ for the case of three equivalent bands with $g_{12}<0$ 
in Fig.19.  $x_2$ vanishes at $|\gamma|=\rho/3$ and is large for large 
$|\gamma|$.

\begin{figure}[htbp]
\begin{center}
\includegraphics[width=6.8cm]{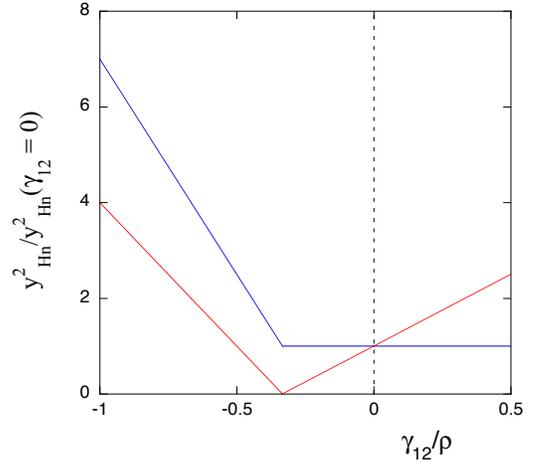}
\caption{$y_{Hn}^2$ ($n=1,2,3$) as a function of $\gamma_{12}/\rho_1$
where we assume an isotropic three-band superconductor and $g_{12}<0$
(the case B in the text).
We adopted that $g_{11}=g_{22}=g_{33}$, $g_{ij}$ ($i\neq j$) are the same, 
and $N_1(0)=N_2(0)=N_3(0)$.
In this model,  two Higgs states are degenerate and the other one state becomes
a massless mode at $\gamma_{12}/\rho_1=-1/3$.
}
\label{mH3gap4}
\end{center}
\end{figure}

\subsection{Discussion}

We have considered the Higgs modes in multi-gap superconductors.
Since the upper critical field $H_{c2}$ is proportional to $y_H^2$,
there is a possibility that we have large $H_{c2}$ in a multi-gap
superconductor such as iron-based superconductors by tuning
interaction parameters.
The eigenvalue $y_H$ increases as $|{\rm det}g|$ decreases.
In a two-gap superconductor, we have two solutions for the gap equation 
and one solution with higher $T_c$ is realized.
The other solution with low $T_c$ is expected to be less important.
When the coherence length of low-$T_c$ solution is shorter than that of the
high-$T_c$ solution, we expect that the low-$T_c$ solution plays
a role in determining the critical field.  As a result the upper critical
field may be larger.
This gives a possibility of high upper critical field in a multi-gap
superconductor.

\section{Summary}

We have examined the property of Green's functions of the Nambu-Goldstone
and Higgs modes in a superconductor. 
In an $N$-gap superconductor, there are $N$ Nambu-Goldstone modes.
We have, however, one gapless mode and $N-1$ massive modes in the
presence of interband BCS couplings.
The NG mode Green function $D(\omega,{\bf q})$ for small $\omega$ and
${\bf q}$ is given as
$D\propto 1/(\omega^2-\omega({\bf q})^2)$ with the dispersion 
$\omega({\bf q})=v_{NG}|{\bf q}|$ where
$v_{NG}=(1/\sqrt{3})\sqrt{ (N_1v_{F1}^2+\cdots+N_Nv_{FN}^2)/(N_1+\cdots+N_N) }$.
An analytic property of the NG Green function $D(\omega,{\bf q})$ is
dependent on $\omega$.
One gapless mode remains gapless in the presence of intraband scattering due
to non-magnetic and magnetic impurities, which was shown on the basis of
the Ward-Takahashi identity.
In a multiband superconductor, massive modes due to interband couplings
$g_{nm}$ become gapless again in a region with time-reversal symmetry
breaking.

The Higgs Green function was also examined.
The time-dependent part of the Higgs action is dependent on the
temperature; it is given by $(\partial_t h)^2$ at low temperature,
while it is $h\partial_t h$ near the critical temperature $T_c$.
The Green function $H(\omega,{\bf q})$ of the Higgs mode has a singularity at
$\omega\sim 2\Delta$ given as
$2\Delta/\sqrt{(2\Delta)^2+\frac{1}{3}v_F^2{\bf q}^2-\omega^2}$.
The Higgs Green function has the same singularity as the $\sigma$-boson
Green function in the Gross-Neveu model.
We have shown that
when there are several order parameters, the constant part of
the Higgs action is important and
crucially dependent upon the interband coupling constants $g_{ij}$.
In a multiband superconductor, the eigenvalue of the matrix $M$ of constant 
Higgs potential
can be very large as the interband coupling constant $g_{12}$ increases,
although the other eigenvalues of the other Higgs mode remain constant.
This indicates the possibility of the large upper critical field $H_{c2}$
because of the relation $H_{c2}\propto 1/\xi^2 \propto y_H^2$.
In iron-based superconductors, the extremely  huge $H_{c2}$ has been 
reported for NdFeAsO$_{0.7}$F$_{0.3}$\cite{jar08} and
Ba$_{0.6}$K$_{0.4}$Fe$_2$As$_2$\cite{wan08}.
Our results indicate that the huge $H_{c2}$ may be due to the
multiband effect for the Higgs modes. 

This work was supported in part by Grant-in-Aid from the Ministry
of Education, Culture, Sports, Science and Technology of Japan
(Grants No. 22540381 and No. 17K05559).

\end{document}